\documentclass[10pt,journal,compsoc]{IEEEtran}
\newif\ifpeerreview

\peerreviewfalse

\usepackage[nocompress]{cite}
\usepackage{url}
\usepackage{amsmath,amssymb,graphicx}

\usepackage{lipsum} 

\usepackage[switch]{lineno}

\usepackage{ulem}
\usepackage{siunitx}
\usepackage{tabto}
\usepackage[font=small,skip=0pt,belowskip=-15pt,aboveskip=0pt]{caption}
\usepackage{float}

\usepackage{titlesec}
\titlespacing*{\section}{0pt}{1\baselineskip}{\baselineskip}

\newcommand{\paperID}{16}

\newcommand{\Section}[1]{\vspace{-4pt}\section{#1}\vspace{-4pt}}
\newcommand{\SubSection}[1]{\vspace{-4pt}\subsection{#1}\vspace{-3pt}}

\title{High Resolution, Deep Imaging Using Confocal Time-of-flight Diffuse Optical Tomography}

\author{Yongyi Zhao*, Ankit Raghuram*, Hyun K. Kim, Andreas H. Hielscher,\\Jacob T. Robinson, and Ashok Veeraraghavan%
\IEEEcompsocitemizethanks{\IEEEcompsocthanksitem *Authors contributed equally
\IEEEcompsocthanksitem Y. Zhao, A. Raghuram, J. T. Robinson, and A. Veeraraghavan are with the Department of Electrical and Computer Engineering, Rice University, Houston, TX 77005.\protect\\
E-mail: \{yongyi, ar89, jtrobinson, vashok\}@rice.edu
\IEEEcompsocthanksitem H. K. Kim with the Department of Radiology, Columbia University, New York, NY and the Department of Biomedical Engineering, New York University, New York City, NY 11201.\protect\\
E-mail: hkk2107@cumc.columbia.edu
\IEEEcompsocthanksitem A. H. Hielscher is with the Department of Biomedical Engineering, New York University, New York City, NY 11201\protect\\
E-mail: ahh4614@nyu.edu}%
}
\begin{document}

\IEEEtitleabstractindextext{%
\begin{abstract}
Light scattering by tissue severely limits how deep beneath the surface one can image, and the spatial resolution one can obtain from these images. Diffuse optical tomography (DOT) is one of the most powerful techniques for imaging deep within tissue -- well beyond the conventional $\sim$10-15 mean scattering lengths tolerated by ballistic imaging techniques such as confocal and two-photon microscopy. Unfortunately, existing DOT systems are limited, achieving only centimeter-scale resolution. Furthermore, they suffer from slow acquisition times and slow reconstruction speeds making real-time imaging infeasible. We show that time-of-flight diffuse optical tomography (ToF-DOT) and its confocal variant (CToF-DOT), by exploiting the photon travel time information, allow us to achieve millimeter spatial resolution in the highly scattered diffusion regime ($> 50 $ mean free paths). In addition, we demonstrate two additional innovations: focusing on confocal measurements, and multiplexing the illumination sources allow us to significantly reduce the measurement acquisition time. Finally, we rely on a novel convolutional approximation that allows us to develop a fast reconstruction algorithm, achieving a 100$\times$ speedup in reconstruction time compared to traditional DOT reconstruction techniques. Together, we believe that these technical advances serve as the first step towards real-time, millimeter resolution, deep tissue imaging using DOT.
\end{abstract}

}

\maketitle
\thispagestyle{empty}

\begin{figure}[t]
\centering\includegraphics[width=1\linewidth]{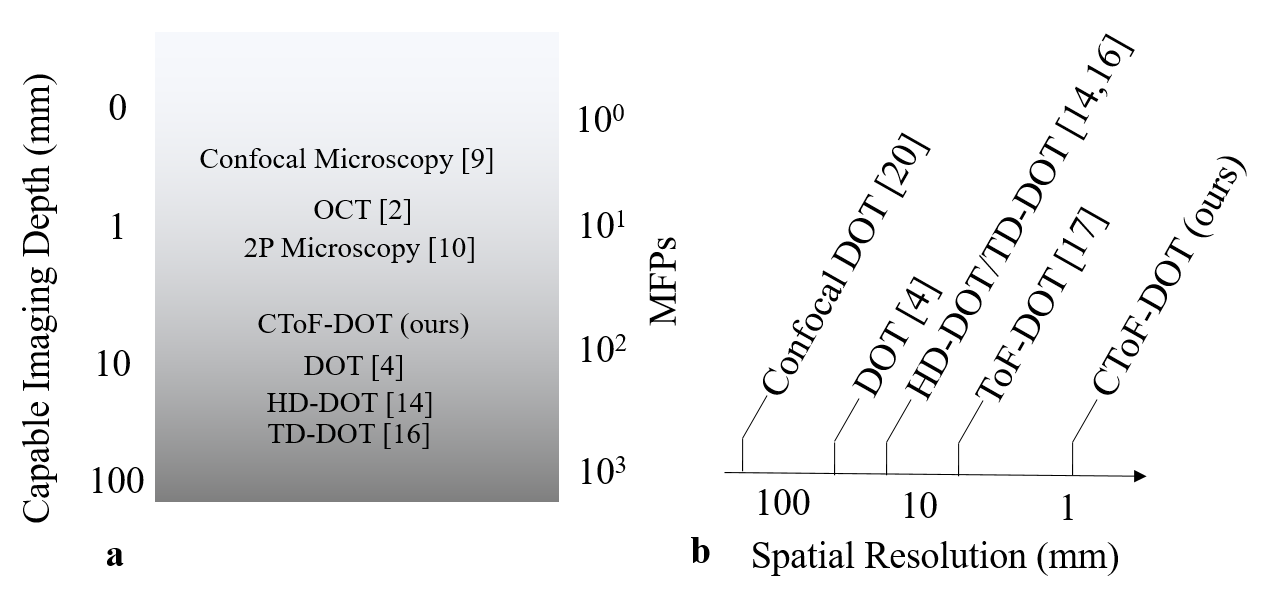}
\caption{\textbf{Imaging depth and spatial resolution of DOT techniques.} (a) Approximate imaging depth of optical imaging techniques. Ballistic imaging techniques such as OCT, confocal microscopy, and 2P microscopy cannot image past $\sim$15 mean free paths (MFPs). DOT approaches can achieve 10s-100s of MFPs (b) Approximate spatial resolution of different DOT techniques. Our technique is the only method to demonstrate 1 \si{\milli\meter} spatial resolution.}
\label{fig:dot-comparison}
\end{figure}

\IEEEraisesectionheading{
  \Section{Introduction}\label{sec:introduction}
}
\IEEEPARstart{L}{ight} scattering by tissue is the primary challenge limiting our ability to exploit non-ionizing, optical radiation in the 400-1000 \si{\nano\meter} wavelength range, to perform high-resolution structural or functional imaging, deep inside the human body.
Most existing techniques, including confocal microscopy, two-photon (2P) microscopy, and optical coherence tomography (OCT), exploit only the ballistic (or single-scattered) photons and can only be used to image within the ballistic regime ($<15$ mean scattering lengths deep) \cite{Pediredla2016,oh2019skin}. This limits imaging to approximately the top 1-2 millimeters of tissue surface (as mean scattering lengths in tissue is $\approx 50-150$ \si{\micro\meter} range \cite{Pediredla2016, Bevilacqua1999}) as seen in Fig. \ref{fig:dot-comparison}a. Many applications (both clinical and scientific) require imaging at much higher depths of penetration than can be achieved by remaining within the ballistic regime.

Diffuse optical tomography (DOT) \cite{boas2001imaging} has emerged as one of the most promising techniques (another being photo-acoustic tomography \cite{Xia2014}) for high-resolution imaging deep within tissue, in the diffusion regime (i.e., $>50$ mean scattering lengths). The idea in DOT is that even in the diffusive regime, where light paths are highly random, there are statistically predictable structures in its distribution in space, and this regularity can be exploited if sufficient diversity of measurements are obtained.
DOT uses an array of sources and detectors placed over the imaging volume -- and the light transport data acquired between each source-detector pair provides the required measurement diversity.

\SubSection{Challenges, Key Ideas, Impacts, and Limitations}

\textbf{Challenges.} In spite of its promise, DOT systems today remain severely limited.
Firstly, existing DOT systems provide low spatial resolution. 
Most are limited to \si{\centi\meter}-scale spatial resolutions because of a combination of factors including lack of sufficient measurement diversity, modeling inaccuracies, and low SNR measurements (Fig. \ref{fig:dot-comparison}b).
Second, the sequential nature of DOT measurement process introduces a trade-off between SNR and capture time, further limiting resolution (and quality) when it comes to imaging dynamics.
Third, DOT reconstruction algorithms have to contend with solving large-scale optimization problems with potentially millions of variables and therefore tend to be quite slow, precluding real-time performance.
Our goal, in this paper, is to directly address these limitations.

\noindent\textbf{Key Ideas.} Our approach leverages three key ideas.

\textit{Key Idea 1 - Increased measurement diversity provided by transients.}
The primary cause of reduced spatial resolution is understood to be the limited measurement diversity.
Increasing the number of source-detector pairs improves spatial resolution but this tends to saturate beyond a point.
It becomes essential to enhance the diversity of measurements by adding additional dimensions.
Time of travel between source and detector may be a promising additional dimension that is significantly beneficial since many of the surface scattered background photons tend to have a significantly shorter travel time than most of the deep penetrating signal photons that interact with the tissue of interest \cite{puszka2013time}.
We demonstrate that exploiting this additional transient dimension (by capturing transient histograms between every source-detector pair), provides a sufficient increase in measurement diversity to obtain \si{\milli\meter} spatial resolution even in the diffusive regime.

\textit{Key Idea 2 - Reduced capture time through multiplexed measurements.}
DOT measurements are typically acquired sequentially and this establishes a trade-off between capture time and SNR.
We propose that multiplexed acquisition, wherein multiple light sources are 'on' simultaneously, improves measurement SNR.
With a reconstruction algorithm that can de-multiplex these measurements, we show that source multiplexing can provide a $4\times$-$10\times$ reduction in capture time compared to traditional sequential DOT.

\textit{Key Idea 3 - Real-time reconstruction using a novel convolutional approximation.} 
Traditional DOT reconstruction algorithms are already computationally intensive --- and with the $\sim$100$ \times$ increase in measurement dimensionality imposed by capturing transient information, this burden is severely exacerbated precluding any hope for near real-time reconstruction performance. 
We propose a novel convolutional approximation for multiplexed (and non-multiplexed), confocal time-of-flight diffuse optical tomography and utilize this approximation to develop a fast, real-time reconstruction algorithm (which is a $100\times$-$1000\times$ speedup).

\noindent\textbf{Outcomes and Potential Impacts.}
The primary outcome that we are able to demonstrate is that we show millimeter spatial resolution in the diffusive regime ($>50$ mean scattering lengths).
This, in itself, opens up a variety of new clinical and scientific imaging applications.
In particular, we believe that non-invasive brain imaging (both structural and functional) is a critical application domain.
As skull severely attenuates acoustic waves making through-skull photo-acoustic tomography difficult \cite{Nie2012}, DOT already is the predominant technology for this application.
Improving the achievable spatial resolution will provide us better specificity potentially allowing us to image columnar fields in the brain for the first time.
The secondary outcome is the first demonstration of a real-time reconstruction algorithm for time-of-flight DOT. In addition, we also show that multiplexing can significantly reduce capture-time in DOT.
Finally, we develop two different versions of the algorithm for both fluorescence and absorption imaging and demonstrate real results for both modes - expanding the potential scope of applications.

\noindent\textbf{Limitations.}
All our current demonstrations are in tissue samples and phantoms (both fluorescence and absorption). 
We are actively working towards demonstrating the feasibility in real biological tissue, as we realize that there are additional challenges such as reduced fluorescence/absorption contrast, increased biological noise, and motion (especially when imaging \textit{in vivo}) that we might need to address before the technology can reach its promised potential.

Our current prototype is sub-optimal in many respects.
While traditional DOT systems have a wearable form-factor, our laboratory prototype uses benchtop optics, with a scanned laser head and a single detector being scanned to mimic a detector array.
While compact systems with a similar wearable form-factor to existing DOT systems are indeed possible, this requires fabrication of an array of SPAD detectors and corresponding laser diodes, with a common shared clock -- something that is beyond the scope of this paper.
Since our benchtop prototype scans a single pixel to emulate a detector array, the total scan time of our system is increased by a factor that is proportional to the total number of detectors being emulated (typically $100\times$-$400\times$  in our results). 
    This along with scanning inefficiencies mean that the scan time in all our results is in the several seconds to minutes range, precluding any ability to show real-world dynamics in our real results.
We are working towards realizing a compact, fabricated, prototype for a wearable, brain imaging system and are hopeful that we can demonstrate that system in action in about a year.  

\Section{Related Work}
\noindent\textbf{Imaging within the ballistic regime.} The fraction of photons that enter a tissue and remains ballistic decreases exponentially with the thickness of the tissue being imaged.
Even after just 3 mean scattering lengths, the fraction of photons that are ballistic become 1 in 20 according to the Beer-Lambert Law \cite{Wang2007}.
As a consequence, even at these depths, techniques such as direct imaging, brightfield imaging, or fluorescence imaging that do not actively filter out the scattered photons get overwhelmed by the background from these multiply-scattered photons reducing the imaging contrast to below the sensor sensitivity thresholds \cite{Pediredla2016}.

 Beyond this depth active means of rejecting the multiply-scattered photons are needed.
 Confocal microscopy uses a set of matched pinholes to reject a large fraction of the scattered light, and typically extends imaging to about 6 mean scattering lengths (1 in 400 photons are ballistic) \cite{kempe1996comparative}.
 Multi-photon microscopy techniques including 2P microscopy, rely on the non-linear excitation process to confine fluorescent emission, and these techniques may allow imaging to be performed as deep as 16 mean scattering lengths (1 in $\sim$8.8 million photons are ballistic) \cite{sergeeva2010scattering}.

Going beyond this ballistic regime of operation is inherently challenging because of the low fraction of photons that remain ballistic.
At 20 mean scattering lengths, 1 in $\sim$480 million photons remains ballistic.
Going beyond as you encroach into the diffusive regime ($\sim$50 mean scattering lengths and beyond), techniques that rely exclusively on ballistic (or single-scattered) photons are completely infeasible as less than 1 in $5.2\times10^{21}$ photons are ballistic.

\noindent\textbf{Beyond the ballistic regime.}
As you move beyond the ballistic regime, the fraction of ballistic photons is so small that relying on them exclusively is insufficient.
Therefore, it becomes imperative, to find ways to model the localization (even if it is only partial) of the scattered photons and exploit these scattered photons as well.

\noindent\textbf{Diffuse Optical Tomography (DOT).}
DOT originated in the 1990s as a way of detecting absorption changes in medical imaging applications \cite{puszka2013time}. 
Traditional DOT systems utilize an array of near-infrared, continuous-wave (CW) light sources illuminating the tissue, resulting in multiply scattered photons that arrive at an array of detectors \cite{boas2001imaging, Liu2020}. 
Models of photon propagation physics could then infer local absorption and scattering properties within the tissue from the measurements captured by the detectors. 
Early applications of DOT included imaging tumors for breast cancer and monitoring brain bleeds for infants \cite{boas2001imaging,puszka2013time}.
Transmittance measurements of these geometries provided absorption information on the whole volume of interest. 
However, the adult brain and internal organs must be imaged in reflection mode due to the strong scattering and absorption properties, or limited access to the tissue of interest \cite{puszka2013time}. 
In the rest of the paper, we will refer to continuous-wave DOT as DOT for simplicity.

Recent advances in DOT have been focused on algorithmic improvements resulting in higher spatial resolution \cite{Liu2020} and development of wearable devices \cite{zhao2017review,frijia2020functional,wheelock2019high}. 
The most significant drawback of DOT is depth sensitivity. 
For deeper penetration in reflection mode, source and detector separations must be farther apart, reducing the SNR of the measurements \cite{Liu2020}. Frequency- and time-domain DOT have been developed to counteract these problems. 
While both frequency- and time-domain can capture the same information, time-domain DOT (TD-DOT) can make measurements faster, albeit with more expensive hardware \cite{gibson2005recent}.  

\noindent\textbf{ToF-DOT.} ToF-DOT (or TD-DOT) uses a high-power, narrow pulse-width laser and a fast-gated detector to capture transient light transport data\cite{pifferi2016new,Lyons2019}. 
These transients contain photon arrival time information for each source-detector pair, providing an additional dimension of information to improve depth sensitivity \cite{pifferi2016new}. 
The emergence of single-photon avalanche diodes (SPADs) in recent years coupled with on-chip time-correlated single photon counting (TCSPC) electronics has allowed for fast-gated, large dynamic range, \si{\pico\second} resolution transient measurements in reasonable acquisition times, making ToF-DOT a promising technology to explore \cite{puszka2013time}.
In addition, hardware improvements are making wearable ToF-DOT systems feasible, and there is some early work towards that direction \cite{farina2017time,di2017miniaturized}.

\noindent\textbf{Comparison with confocal diffuse tomography \cite{Lindell2020}.} Recently, Lindell et al. proposed a confocal diffuse tomography system, which is a very similar system to CToF-DOT \cite{Lindell2020}. While the predominant application of DOT and ToF-DOT has been deep-tissue imaging (especially breast cancer and through-skull imaging), their technology focuses on other applications such as imaging through thick scatterers, and imaging around a corner.
Similar to CToF-DOT, their work also demonstrated that a 3D image can be acquired through thick scattering media. Additionally, both \cite{Lindell2020} and CToF-DOT apply a linear forward model. However, there are also several notable differences. Firstly, the CToF-DOT forward model assumes the target features are embedded within a scattering media, while the forward model in \cite{Lindell2020} assumes the targets are at some standoff distance from the scattering media. Secondly, CToF-DOT must contend with greater background noise due to the physical proximity between the target features and the tissue phantom surface. Thirdly, CToF-DOT can handle multiple imaging modalities, including absorption or fluorescence targets, and can be extended to scattering. These differences are based on the separate application spaces: CToF-DOT for non-invasive neuroimaging and \cite{Lindell2020} for LiDAR and NLOS imaging.

The principal limitation of DOT and ToF-DOT remains the limited spatial resolution provided by this approach.
Existing DOT and ToF-DOT systems \cite{puszka2015spatial,boas2001imaging}, have only been able to demonstrate \si{\centi\meter}-scale spatial resolution.

\noindent\textbf{Reconstruction Algorithms.} DOT reconstruction approaches have traditionally focused on iteratively solving approximations of the Radiative Transfer Equation. 
Analytical solutions to the radiative transfer equation only exist for the most simple examples, and for any scenario approaching real-world complexity, numerical techniques are the only alternative.
This numerical process for solving the radiative transfer equation is computationally challenging, resulting in reconstruction algorithms that take hours to converge \cite{Kim2008}. 
Fortunately, photon propagation can be reformulated as a linear system using the Born approximation. 
Then solving for the optical properties can just be a linear inverse problem, thereby speeding up reconstruction algorithms \cite{boas2001simultaneous,tarvainen2010corrections}.  
However, these algorithms still require the storage of an extremely large sensitivity matrix and therefore suffer from increases in the dimensionality of the measurements. 
In summary, even the fast reconstruction techniques such as \cite{naser2015time,mozumder2020time} typically end up taking several 10s of seconds to minutes per iteration.

This computational challenge is further exacerbated in ToF-DOT where the inclusion of transient information adds an additional dimension to the problem.
As a result, naive attempts at high-resolution reconstruction for such ToF-DOT systems, by directly incorporating time of travel information within the existing DOT algorithms, can lead to far greater reconstruction times as a result of dimensionality.
As a result, there exist no real-time (or near real-time) reconstruction algorithms for ToF-DOT systems.

Efficiently and accurately estimating optical properties also persists in inverse rendering in the Computer Vision/Graphics discipline. \cite{gkioulekas2013inverse, gkioulekas2016evaluation} developed efficient inverse scattering methods for inverting the radiative transfer problem while \cite{nimier2019mitsuba,nimier2020radiative,azinovic2019inverse} used differentiable rendering approaches to resulting in faster light transport simulations. \cite{che2018inverse} replaced the forward operator with a trained neural network for fast and accurate renderings. Inverse rendering approaches have shown success in atmospheric tomography \cite{levis2015airborne,levis2017multiple} and imaging through fog \cite{satat2018towards}, the latter of which takes advantage of time-of-flight information to reconstruct object and object depths. Ultimately, in CToF-DOT, we took an alternative approach of reducing the size of the Jacobian and applying a convolutional approximation and found that it allowed us to achieve reasonable memory and computational efficiency in our application.

\Section{ToF-DOT}
\begin{figure}[t]
\begin{center}
   \includegraphics[width=1\linewidth]{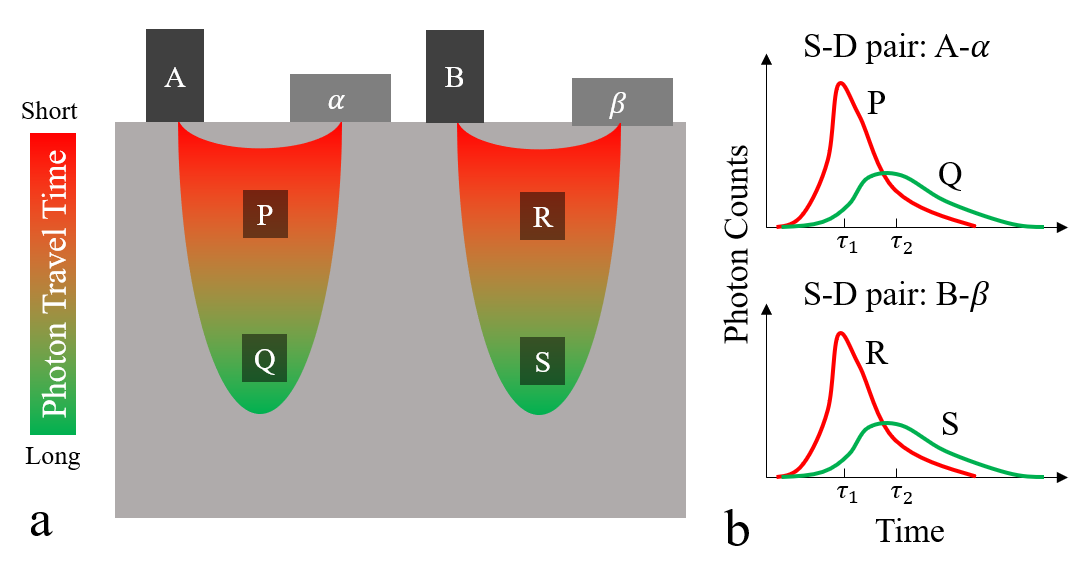}
\end{center}
   \caption{\textbf{ToF-DOT concept.} (a) Photon trajectories for 2 source-detector pairs. A and B are sources, $\alpha$ and $\beta$ are detectors, and P, Q, R, and S are voxels of interest. Source-detector pair A-$\alpha$ is more sensitive to P and Q and source-detector pair B-$\beta$ is more sensitive to R and S. (b) Photon arrival times passing through specific voxels associated with source-detector pair A-$\alpha$ and B-$\beta$. Photons passing through voxels closer to the surface (P and R) tend to arrive earlier than photons passing through voxels deeper inside (Q and S).}
\label{fig:03_concept}
\end{figure}

\begin{figure}[t]
\begin{center}
   \includegraphics[width=1.0\linewidth]{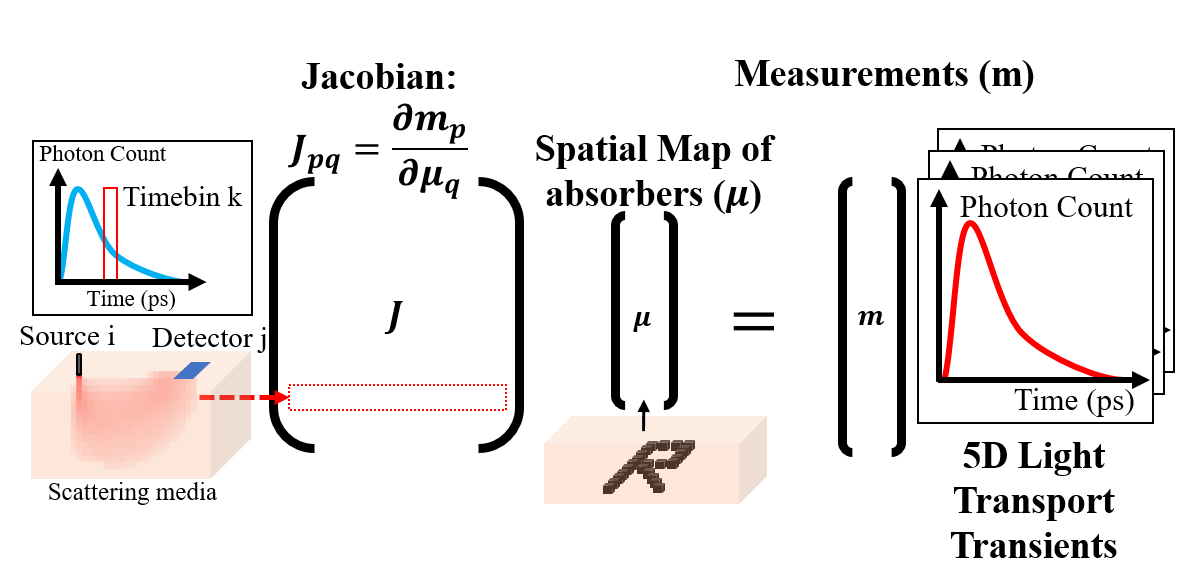}
\end{center}
\caption{\textbf{Overview of DOT forward model.} In the linear forward model, the target scene ($\mu$) is mapped to a set of measurements ($m$) by the Jacobian matrix ($J$)}
\label{fig:02_fwd_model_overview}
\end{figure}

A traditional DOT system consists of an array of light sources and an array of detectors placed on top of the imaging volume.
Shown in Fig. \ref{fig:03_concept}(a) is a statistical distribution of light transport paths between two different source-detector pairs. 
Intuitively, each of these intensity light transport measurements can be thought of containing weighted information about the attenuation (absorption) or emission (fluorescence) from the voxels in the imaging volume. 
The weights themselves can be intuitively thought of as being approximately proportional to the likelihood that light paths for that source-detector pair traverse through that particular voxel.
In the example shown in Fig. \ref{fig:03_concept}(a), intensity light transport measurement between source $A$ and detector $\alpha$ contains more information about voxels $P$ and $Q$, while intensity light transport measurement between source $B$ and detector $\beta$ contains more information about voxels $R$ and $S$.

In ToF-DOT, the light sources are typically ultra-short pulsed sources, and the detectors measure transient (or time of travel) information in addition to the intensity.
Thus, for each source-detector pair, the transient light transport information is recorded.
Shown in Fig. \ref{fig:03_concept}(b), is a statistical distribution of light transport paths between a source-detector pair, where the time of travel of these paths are also color-coded.
Clearly, the original intuition behind DOT holds true.
But in addition to that, we notice that photons with different travel times pass through very different locations within the imaging volume, providing us additional information about spatial localization.
In the example shown in Fig. \ref{fig:03_concept}(b), transient light transport measurement  with a travel time of $\approx \tau_1$, contains more information about voxel $R$, while transient light transport measurement with a travel time of $\approx \tau_2$, contains more information about voxel $S$.

The primary advantage of ToF-DOT is that this additional transient information has the potential to significantly improve spatial resolution in the reconstructions.

\SubSection{Transient Light Transport: Forward Model}
The propagation of light though a scattering media is well-modeled using the radiative transfer equation (RTE) \cite{Wang2007}:
\begin{equation}
\begin{split}
    \frac{\partial L(\Vec{r}, \hat{s}, t)/c}{\partial t}=-\hat{s}\cdot \nabla L(\Vec{r}, \hat{s}, t)-\mu_t L(\Vec{r}, \hat{s}, t) + \\
    \mu_s\int_{4\pi} L(\Vec{r}, \hat{s}', t)P(\hat{s}'\cdot \hat{s})d\Omega'+S(\Vec{r}, \hat{s}, t) \label{rte}
\end{split}
\end{equation}
Where $L(\Vec{r}, \hat{s}, t)$ is the radiance at a particular position $\Vec{r}$, solid angle $\hat{s}$, and time $t$; $P(\cdot)$ is the phase function, which describes the scattering angle; $S(\cdot)$ is the source term; and $\mu_t$ is the extinction coefficient. 

To create a more computationally tractable problem, we wish to linearize the RTE. This linearization assumes that the absorbers affect the path of only a few photons, creating a negligible change in the photon distribution outside of the absorber region. This superposition assumption, which is essential to the linearization, only breaks down in the presence of large absorbers altering the photon path distribution. A more rigorous derivation of this can be found in references \cite{Kak1988, Yao2018}. As a result, the RTE can be reformulated as a linear equation by considering the differential measurements \cite{Yao2018}:
\begin{equation}
m = J\mu,
\label{linear-eqn} 
\end{equation}
where $\mu$ represents the spatially varying material properties within the imaging volume, $m$ is the transient light transport measurements acquired by ToF-DOT, and $J$ is the Jacobian or sensitivity matrix.
Going back to our intuition, the Jacobian, $J$, nominally represents the weights of each voxel in the volume to each measurement. An overview of this linear model is shown in Fig. \ref{fig:02_fwd_model_overview}.

Human tissue and most other biological tissues (including the skull for example) are predominately scattering and have little absorption. So it is reasonable to assume that native tissue absorption can be ignored. In addition, tissue optical properties are fairly uniform, with some significant heterogeneities that correspond to physiologically important variations. So, these properties are modeled as the summation of a spatially homogeneous background material coefficient ($\mu_0$) and a foreground, spatially varying material coefficient that is typically the imaging property of interest ($\mu$). In Equation \eqref{linear-eqn}, $\mu$ represents the spatial distribution of these heterogeneities in the scattering media. These heterogeneities can be fluorophores (emission signal) or optical absorbers (absorption). 
In a biological context, they can represent features of interest such as tumors, vasculature, or regions of biological activity. 

If the imaging volume is discretized into $N_{voxels} = L \times W \times H$ voxels, then $\mu$ is a vector of length $N_{voxels}$, that represents the tissue heterogeneities. Let us assume a ToF-DOT system consists of $N_{s}$ sources of light, $N_{d}$ detectors, wherein each detector measures a transient that is then binned into one of $N_{t}$ time bins. In this case, the set of all measurements can be represented as a vector $m$ of length $N_{meas}= N_s \times N_d \times N_t $. The two quantities $\mu$ and $m$ are related by the Jacobian, $J$, which is a matrix of dimension $N_{meas} \times N_{voxels}$, where $J_{pq} = \frac{\partial m_p}{\partial \mu_q}$. Each entry of the Jacobian, $J_{pq}$ defines the sensitivity of measurement $m_p$ to a corresponding heterogeneity $\mu_q$. 

\SubSection{Computation of the Sensitivity Matrix}
In order to leverage the linear approximation in Equation \eqref{linear-eqn}, one needs to first obtain an accurate estimate of the sensitivity matrix $J$.
In practice, there are two potential ways to estimate the sensitivity matrix: (a) fast analytical approximation, or (b) accurate but slow Monte-Carlo simulation.
Note that in either case, the computation of the sensitivity matrix is a one-time process for any application and need not be real-time. 

In cases for which the Jacobian becomes too large or computationally expensive to compute, certain inverse rendering applications rely on gradient computations and skip the computation of the Jacobian \cite{nimier2020radiative, gkioulekas2016evaluation, levis2015airborne, levis2017multiple}. One example of this adjoint radiative transfer methods. However, in our approach, we opted to compute the Jacobian as we are not limited in Jacobian compute time.

\noindent\textbf{Analytical approximation.}
Using the diffusion approximation, we can derive a closed-form approximation to the RTE \cite{Liu2020, Wang2007, Hyde2002}. 
According to \cite{Liu2020}, we can derive this equation using the Born Approximation: 
\begin{equation}
    m(\vec{r}_d, \vec{r}_s)=\int_v \bigg(\Phi_0(\vec{r}_v-\vec{r}_s)R(\vec{r}_d-\vec{r}_v)\bigg)\mu(\vec{r}_v)d\vec{r}_v
    \label{eq:fwd_eqn_cw}
\end{equation}
Where $\Vec{r}_s$, $\Vec{r}_d$, $\Vec{r}_v$ are the positions of source $s$, detector $d$, and voxel $v$ respectively; $m(\vec{r}_d, \vec{r}_s)$ is the measurement as a function of source-detector position; $\mu(\vec{r}_v)$ is the spatial distribution of optical properties, i.e. the image of interest; $\Phi_0(\vec{r}_v-\vec{r}_s)$ and $R(\vec{r}_d-\vec{r}_v)$ are the fluence rate and diffuse reflectance terms. This product is the Jacobian:
\begin{equation}
    J(\vec{r}_s, \vec{r}_d, \vec{r}_v) = \Phi_0(\vec{r}_v-\vec{r}_s)R(\vec{r}_d-\vec{r}_v)
    \label{eqn:cw-jacobian}
\end{equation}
Equation \eqref{eqn:cw-jacobian} can be adapted to time of flight measurements by calculating the time-domain convolution of the Green's function and reflectance rather than the direct product as shown by Hyde \textit{et al.} \cite{Hyde2002}:
\begin{equation}
J(\vec{r}_s, \vec{r}_d,\vec{r}_v, t)=\Phi_0(\vec{r}_v-\vec{r}_s, t)\circledast_t R(\vec{r}_d-\vec{r}_v, t) \label{closed-form-eqn}
\end{equation}
In the time-domain, Equation \eqref{eq:fwd_eqn_cw} becomes:
\begin{equation}
\begin{split}
    m(\vec{r}_d, \vec{r}_s, t)
    &=\int_v J(\vec{r}_s, \vec{r}_d, \vec{r}_v, t)\mu(\vec{r}_v)d\vec{r}_v\label{eq:fwd_eqn_td}
\end{split}
\end{equation}

Because this expression only requires a 1D convolution, it can be used to quickly calculate the Jacobian matrix. 
However, this expression can only be applied to simple scene geometries, such as a single homogeneous slab, and assumes that the scene is highly scattering \cite{Wang2007}. As a consequence, the approximation is not appropriate in many situations such as (a) near the surface, where we are not yet in the diffuse regime, (b) inhomogeneous tissue, or (c) layered tissue that contain low-scattering regions (such as skull, cerebrospinal fluid, and brain) \cite{Kim2017}. 

\noindent\textbf{Monte Carlo (MC) simulations of the forward model.}
While the closed-form expressions can be calculated efficiently, they can be limited by the prior assumptions of a highly scattering media and slab geometry. Since our long-term goal is to tackle brain imaging, we primarily use MC simulations for determining the sensitivity matrix. This technique is widely regarded as the "gold standard" for simulating light propagation \cite{Yao2018, Wang1995}. We apply MC to obtain the Jacobian matrix by estimating solutions to the RTE. In MC, simulated photons are propagated through the imaging volume. Each photon follows a random walk, which is sampled from a distribution that is parameterized by the optical parameters of the scene \cite{Wang1995}. Finally, the aggregate information from many photon samples can be used to estimate the sensitivity matrix. More details on this procedure can be found in Yao \textit{et al.} \cite{Yao2018}.

\SubSection{Reconstruction Algorithm}
The goal of DOT imaging systems is to produce an image reconstruction of the spatial distribution of optical parameters, typically the absorption coefficient, represented by $\mu_a$. This image reconstruction is done using the following optimization setup:
\begin{equation}
min_\mu\lVert m-f(\mu) \rVert + \Lambda(\mu)
\label{general-inverse-eqn} 
\end{equation}
Where, $\mu$ is the spatial distribution of optical parameters; $m$ is the set of collected measurements, which describes the intensity of light incident on the detectors; $f(\cdot)$ is the forward model, which calculates the measured intensity as a function of the optical parameters of the scene and $\Lambda(\mu)$ is an appropriately chosen regularization term.

Using the linear model, the image reconstruction problem can be formulated as a linear inverse problem:
\begin{equation}
    \hat{\mu} = \min_\mu \lVert m - J\mu \rVert_2 + \lambda \lVert \mu \rVert_1,
\end{equation}
where $\lVert m - J\mu \rVert_2$ is the data fidelity term, and $\lVert \mu \rVert_1$ is a regularization term that enforces sparsity in the heterogeneity of the optical properties within the imaging volume, and $\lambda$ is a hyperparameter to tune the sparsity level.
This optimization problem is known to be convex and there are a host of well-understood algorithms that can be used to solve it.
We use the fast iterative shrinkage thresholding algorithm (FISTA) \cite{Beck2009} to solve this optimization since it is fast, has reasonable memory complexity, and has been shown to be accurate (and reaches the global optimal solution).

Even with the use of a fast, iterative algorithm and an implementation on a multi-CPU, multi-core computational system, the algorithm remains too slow to enable real-time applications. As an example, if we consider the reconstruction of a $30 mm \times 30 mm \times 20 mm$ volume at 1 mm voxel size, using a ToF-DOT system that consists of 100 sources and 100 detectors and each transient being binned into 50 different time bins, then the corresponding sensitivity matrix $J$ is of size $500k \times 18k$ and each FISTA iteration on an Intel Xeon machine, with 6 cores takes about 6.3 seconds. Accurate reconstruction may require hundreds of iterations for convergence, meaning that total reconstruction time could be on the order of an hour.

\Section{Confocality and Multiplexing in ToF-DOT}
The computational complexity of current generation ToF-DOT reconstruction algorithms precludes near real-time operation.
A careful study of the computational complexity provides two potential avenues that might facilitate significant improvements in computational speed.

\noindent\textbf{Measurement selection.} 
The computational complexity of solving large-scale linear inverse problems scales between quadratic and cubic in the problem size, based on the kind of algorithms used.
This means that, in practice, while the 10$\times$-100$\times$ increased measurements afforded by ToF-DOT significantly improve the spatial resolution of the reconstruction, it also slows down the reconstruction time by several orders of magnitude compared to traditional DOT algorithms.
One way to combat this is measurement selection, wherein only a select subset of measurements is used in the reconstruction.
To maintain the resolution advantages provided by ToF-DOT, one has to carefully select the measurements so as to ensure that the maximally useful (high SNR, high information gain) measurements are retained.

\noindent\textbf{Faster forward models.}
The key computational step in almost all iterative algorithms (including FISTA) that are intended to solve the optimization problem in Eqn. \eqref{general-inverse-eqn} is the repeated application of the forward operator (or its conjugate or transpose).
In the case of ToF-DOT, this amounts to a matrix multiplication with the corresponding sensitivity matrix (or its transpose) and this matrix multiplication has linear complexity in the number of elements in the matrix (or quadratic in the number of rows/columns).
One key idea that has been in many other applications is if under some restricted regimes of operation, the general linear model can be reduced to a convolutional form, then one could leverage fast implementations of convolutions (that rely on FFTs) to significantly reduce the computational burden.

Here we argue that focusing on confocal ToF-DOT measurements allows us to leverage both these advantages simultaneously, allowing us to achieve, real-time ToF-DOT reconstruction performance.
This would correspond to retaining all the measurements wherein the source and the detector location are the same (or close enough to be modeled as confocal in a real system).

\noindent\textbf{Related work.}
The scanning-time and reconstruction-time challenge in ToF-DOT is not unique to DOT, but rather common across a variety of emerging applications that attempt to utilize the extra temporal dimension offered by transient detectors such as SPADs. These applications include imaging around corners \cite{Velten2012}, non-line-of-sight imaging\cite{Pediredla2019}, and imaging through thick diffusers \cite{Lindell2020} and in all of these examples, the imaging geometry is somewhat similar to ToF-DOT. 
There is an array of sources and detectors that are scanned and transient light-transport measurements are obtained. 
The principal difference between these applications and ToF-DOT is that in these applications, the scattering surface or the thick diffuser acts as an obscurant and the goal is to image objects beyond that obscurant.
In contrast, in ToF-DOT, the goal is to obtain a volumetric image of the optical properties (scattering, absorption, or fluorescence) of the tissue itself. 
Thus the computational model for light propagation in these different applications is quite different.
That said, the symmetry in imaging geometry, between these applications and ToF-DOT, is quite striking.

In all of these applications that use transients, reconstruction algorithms tended to be slow precluding any real-time operation.
Over the last few years, confocality has emerged as a key idea enabling real-time reconstruction in these applications. First, within non-line-of-sight reconstruction, it was shown that restricting the measurements to confocal measurements allows both a reduction in the number of measurements and also enabled a convolutional approximation to the forward model resulting in real-time reconstruction algorithms \cite{Ahn2019, OToole2018}. More recently, similar insight was used to demonstrate near-real-time reconstruction performance for imaging through thick obscurants \cite{Lindell2020}. We are motivated by the success of these techniques and show that this idea, when translated to ToF-DOT, allows us to obtain real-time ToF-DOT reconstructions for estimating 2/3D optical properties of thick tissues.

\SubSection{Confocal ToF-DOT}
Collocated source-detector pair contains the most information as it pertains to deep features; while we provide some intuition for this idea, the following references provide additional detail \cite{pifferi2016new, pifferi2008time, torricelli2005time}. This can be derived based on the idea that the sensitivity matrix represents the likelihood that photons will pass through a particular voxel before reaching the detector. With a closer source-detector separation, this likelihood increases. While this confocality is difficult to implement in DOT due to the increased sensitivity to superficial layers, it can be applied to ToF-DOT because time-domain information mitigates this challenge by rejecting early-arriving photons that are scattered from superficial layers \cite{torricelli2005time, pifferi2008time}. This idea is also reinforced by our results in Fig. \ref{fig:05_singular_vals}. We see that selecting only the measurements from the collocated source-detector pair leads to a more well-conditioned Jacobian matrix than selecting measurements from source-detector pairs of arbitrary separation distance. 

\begin{figure}[t]
\begin{center}
   \includegraphics[width=1\linewidth]{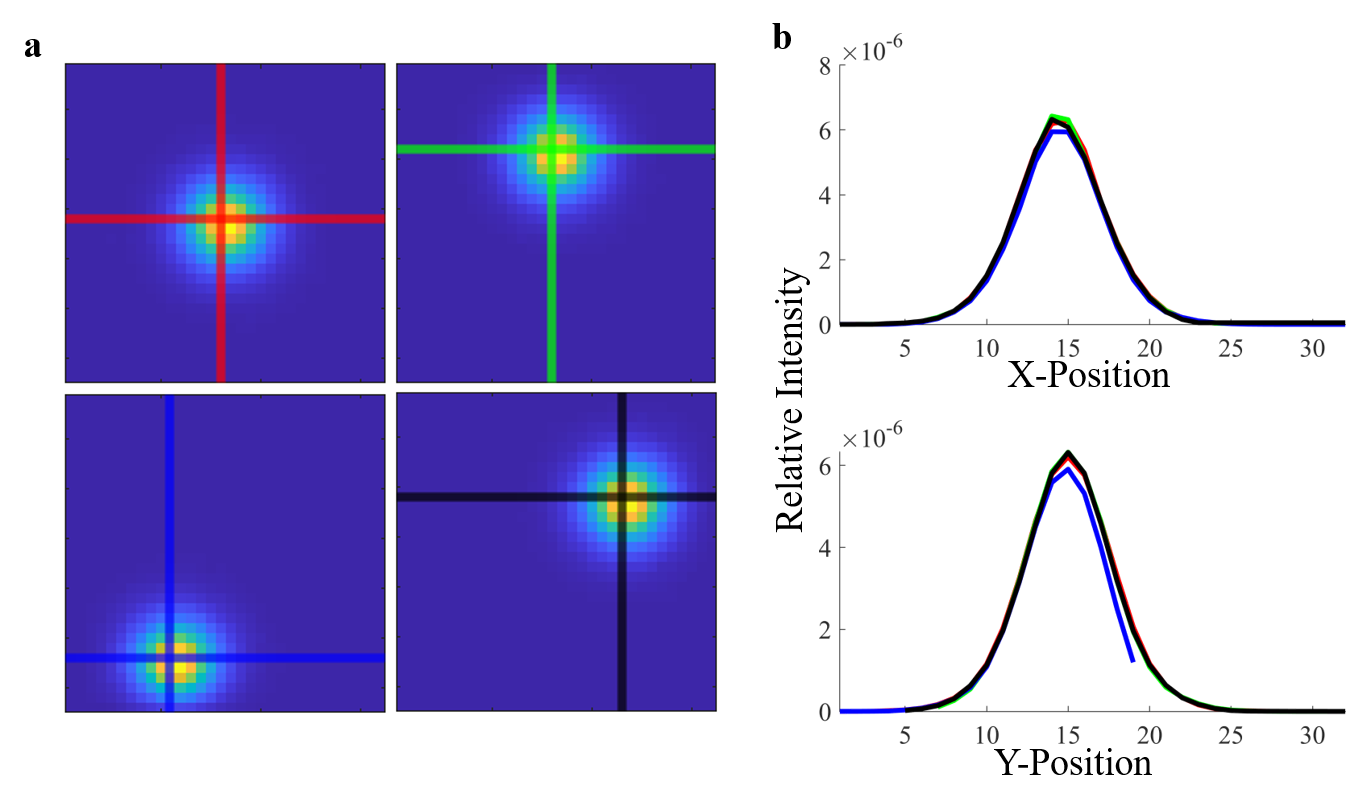}
\end{center}
   \caption{\textbf{Validity of convolutional approximation.} Visualization of (a) rows of Jacobian for different source-detector pair locations, and their corresponding (b) 1D profiles along X and Y directions (colored lines). Note: 1D profiles have been aligned for visualization.}
\label{fig:04_shift_invar}
\end{figure}

\noindent\textbf{Convolutional approximation.}
From equation \eqref{linear-eqn} we see that the forward model for ToF-DOT can be modeled as a linear system.
Fortunately, when restricting our attention to confocal measurements, the linear operator is shift-invariant.
This shift-invariance allows us to develop a convolutional approximation for the confocal ToF-DOT system.

To empirically demonstrate the shift-invariance of the sensitivity matrix (i.e., the matrix is doubly circulant), we use the Monte Carlo simulator to generate different rows of the sensitivity matrix that correspond to point targets at different locations within the volume.
In this simulation, we assume a confocal geometry with features fixed to a specific depth. 
Fig. \ref{fig:04_shift_invar}(a) shows a visualization of $4$ rows of the Jacobian. 
Additionally, from Fig. \ref{fig:04_shift_invar}(b) we see that each blur kernel has the same profile. 
As the feature location is shifted, there is a corresponding shift in the measured output. 
This indicates that the Jacobian is a doubly circulant matrix. 
Therefore, when performing image reconstruction using confocal measurements, we can apply the forward model using a convolutional approximation rather than a matrix-vector product. 
Equation \eqref{linear-eqn} can be substituted with: 
\begin{equation}
    \label{eqn:conv_approx}
    m(x,y,t)=\rho(x,y,t)\circledast \mu(x,y).
\end{equation}
Here $m(x,y,t)$ is the measurement, which is now a function of the collocated source-detector $(x,y)$ and time $t$; $\mu(x,y)$ is the spatially varying material properties, which is now a function of just the lateral positions $(x,y)$; and finally $\rho(x,y,t)$ is the blur kernel. 
The blur kernel $\rho(x,y,t)$ can be determined using either the Monte Carlo simulator or the analytical expressions by calculating (or estimating in the case of Monte Carlo) the measurement for a single feature, i.e. a spatial delta function. 

\noindent\textbf{Extension to 3-dimensional Reconstruction.} The image reconstruction algorithm can also be extended to 3-dimensional (3D) outputs. In this case, the Jacobian matrix is a linear mapping from the 3D set of optical parameters to the collocated measurements. $J:\mu(x_v,y_v,z_v)\rightarrow m(x_{sd},y_{sd})$. Where $(x_v,y_v,z_v)$ and $(x_{sd},y_{sd})$ are the position coordinates of the optical parameters and the collocated source-detector pair, respectively. The same forward model and inverse solver can be applied for solving the image reconstruction problem. To determine each row of the Jacobian, the MC algorithm calculates the sensitivity for every point in the discretized 3D space. After calculating the Jacobian using MC, we apply a linear solver (FISTA) to estimate $\mu$. In our algorithm, we apply a separate regularization parameter for each z-depth. This can be expressed as 
\begin{equation}
    \hat{\mu}(x,y,z) = \min_{\mu} \lVert m - J\mu(x,y,z) \rVert_2 + \lambda(z) \lVert \mu(x,y,z) \rVert_1,
\end{equation}
This layer-wise regularization $\lambda(z)$ is used to counteract the reduced sensitivity at increasing depths. Without it, the output possesses relatively small values for greater depths. Therefore, if the same sparsity regularization is applied to all depths, the values at greater depths would be cast to zero, while reducing the effects of regularization at shallow depths. 

\noindent\textbf{Convolutional Approximation for 3D Reconstruction.} We can also apply a convolutional approximation for reconstructing 3D scenes. Recall that for confocal measurements, the forward model can be rewritten into a convolutional form. From eqn. \ref{eqn:conv_approx}, if the scene is a single plane at depth $z$, we have the following equation:

$$
    m(x,y,t)=\rho_z(x,y,t) \circledast \mu_z(x,y)
$$

Where $\mu_z(x,y)$ and $\rho_z(x,y,t)$ are associated with a fixed depth $z$. Because we are using a linear model, then for a 3-dimensional scene, the forward model becomes:

\begin{equation}
    m(x,y,t)=\sum_z \rho_z(x,y,t) \circledast \mu_z(x,y)
\end{equation}

Where $\rho_z(x,y,t)$ is now a depth-dependent PSF. Thus, when the scene is 3D, the measurement equation can be written as a summation of 2-dimensional convolutions for each depth.

\noindent\textbf{Computational Complexity Analysis.} 
Using the standard forward model, the main bottleneck in solving the inverse problem is the matrix-vector product: $J\mu$. 
This operation scales linearly with the number of sources ($N_s$), number of detectors ($N_d$), time bins($N_t$), and number of voxels ($N_{voxels}$). 
The runtime complexity is $\mathcal{O}(N_s N_d N_t N_{voxels})$. 
The bottleneck for memory usage is the storage of the Jacobian, which is of complexity equal to the matrix size.

This complexity can be significantly reduced in the confocal mode, using a convolutional model. 
Convolution with a size $K\times K$ blur kernel can be efficiently implemented using the fast Fourier transform (FFT). 
Thus, the computational complexity is $\mathcal{O}(N_t K^2 \log (K))$. 
The first improvement is the reduction in the number of measurements from $N_s N_d N_t$ to $N_s N_t$ (in confocal measurements $N_s = N_d$ and only 1 transient measurement is obtained per source location).
The second improvement arises because of the convolutional approximation.

Fig. \ref{fig:06_alg_runtime} shows the significant reduction in computational complexity that is achieved due to the convolutional model imposed on the confocal ToF-DOT measurements. There is a two-orders-of-magnitude speed-up in runtime compared to existing ToF-DOT algorithms \cite{Hyde2002}. Even more remarkable is the resultant confocal ToF-DOT algorithm is over an order of magnitude more efficient than even conventional DOT algorithms \cite{Wang2007} that do not utilize any transient information at all (and result in worse spatial resolution). 

\SubSection{Multiplexed Confocal ToF-DOT}
Traditional DOT systems use point-scanning to capture measurements, which can result in long measurement capture durations precluding the capture of dynamics. 
This challenge is compounded by the fact that DOT systems often require a long exposure duration (even for a single source location), due to the fact that only a minuscule fraction of the incident photons are sensed at the detector -- meaning that the detectors are operating at extremely low photons levels.
We demonstrate that source multiplexing can be used to potentially address both these challenges simultaneously.

\noindent\textbf{Multiplexing sources far away.} 
Typical DOT and ToF-DOT systems have a field of view of the order of $5-10$ cm on a side to image through the skull.
Detectors and sources are typically placed on an array (anywhere from $10 \times 10$ to $25 \times 25$ arrays) with a spacing of a few \si{\milli\meter} to a \si{\centi\meter} between array elements.
When a source is 'on', all detectors are measuring the corresponding light transport transients, but the detectors that are far away typically (i.e., with safe illumination power and within reasonable exposure durations) get little to no photons making their measurements useless.
In practice, the photon signal dies exponentially with separation distance and after about a $2-3$ cm separation there are typically very few photons measured.

This means that one can safely assume that there is no cross-talk between measurements even if multiple sources are kept 'on' simultaneously, as long as we can ensure sufficient separation between the sources.
For each detector measurement, we can allocate the entire transient measurement to the closest source (note that this is only possible when we can ensure that sources that are simultaneously 'on' are sufficiently far away).
In our prototype system with a FOV of about $5$ cm, this means that we can multiplex up to 4 sources simultaneously without any cross-talk.
This allows us to get a 4$\times$ improvement in total capture time, while it does not affect the SNR of the individual measurements.

\noindent\textbf{Multiplexing sources with cross-talk.}
Even in the presence of measurement cross-talk, one can obtain multiplexing gains \cite{Cossairt2013, sankaranarayanan2018hadamard, Schechner2007, Mitra2014}. In this case, the benefits of multiplexing are primarily derived from reducing the effects of signal-independent noise. While SPADs do not suffer from read noise, they are impacted by dark count noise. In simulation, we can decouple the benefits of multiplexing sources far away from multiplexing sources with cross-talk. This is achieved by simulating the effects of background noise, and dark counts, separately. The results of this are shown in supplementary fig. 1.

\noindent\textbf{Composite reconstruction algorithm.}
In the presence of source multiplexing the new measurements $y$ become multiplexed versions of the old measurements $m$ -- wherein $y$ and $m$ are related by the multiplexing matrix $S$ as $y=Sm = S J \mu$.
The combined optimization problem to be solved becomes 
\begin{equation}
    \mu = \min_\mu \lVert y - S J \mu \rVert_2 + \lVert \mu \rVert_1,
\end{equation}
where $J \mu$ can be further efficiently implemented within each iteration using the convolutional approximation. 
As before, we use the fast iterative shrinkage thresholding algorithm (FISTA) \cite{Beck2009} to solve this optimization problem. 

\Section{Materials and Methods}
\noindent\textbf{Simulation setup.}
We use an in-house Monte Carlo simulator for generating measurements and the Jacobian matrices needed for both simulated and experimental results. 
In these simulations, we assume a homogeneous scattering slab. The slab possesses a thickness of 6.5 mm and spans an area of a few centimeters squared. For 2D image reconstructions, we assume the features are restricted to a 2-dimensional plane on the backside of the homogeneous slab; while for 3D reconstruction, we assume a 3-dimensional voxel array.
Our implementation is based on the standard Monte Carlo for scattering samples and closely follows the details in \cite{Wang1995, Yao2018}. 
The Monte Carlo simulations are run on GPUs (Nvidia RTX 2080 Ti). 
Obtaining the Jacobian through analytical expressions was performed on CPU (Intel Xeon 3.30 GHz). 
Finally, our simulator can operate in both fluorescence and absorption imaging modes. 
The details of extending absorption-based Monte Carlo to fluorescence mode are described by Liebert \textit{et al.} and Chen \textit{et al.} \cite{Liebert2008, Chen2011} and we follow these to adapt our implementations as well. In general, to model fluorescence, it is necessary to incorporate the different optical properties between the excitation and emission wavelengths, account for fluorescence lifetime by convolving the simulated TPSF with an exponential parameterized by the fluorescence lifetime, and factor in attenuation of the background signal if an emission filter is present.

\begin{figure}[t]
\begin{center}
   \includegraphics[width=1.0\linewidth]{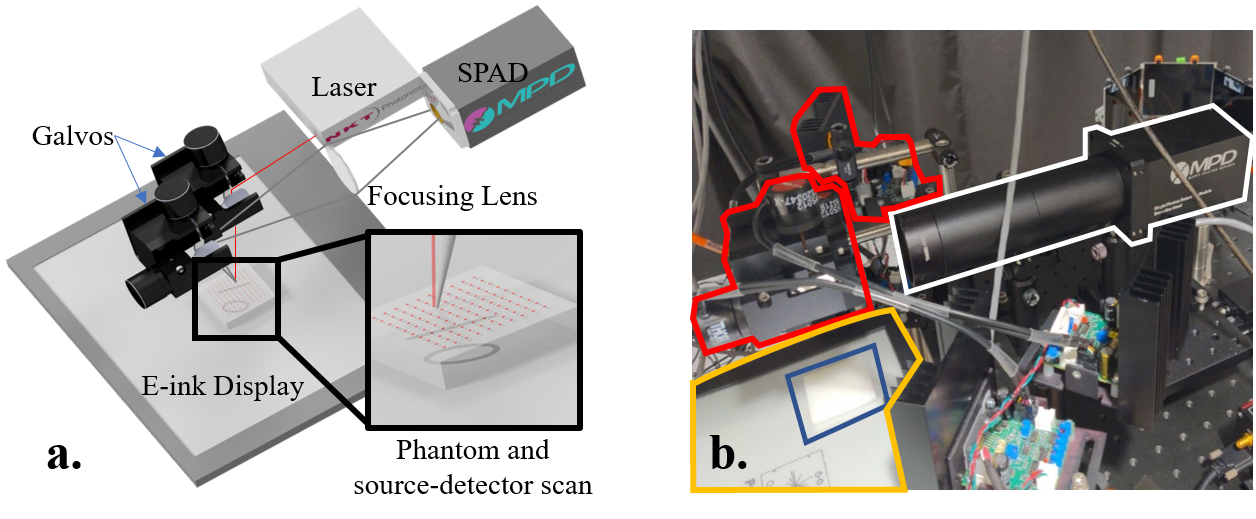}
\end{center}
\caption{\textbf{Experimental setup to test CToF-DOT.} (a) Rendering of our experimental setup showing a scanning laser beam and single pixel detector. (b) An image of the physical setup, with the SPAD (white), galvo mirrors (red), E-ink display (orange) and tissue phantom (blue).}
\label{fig:exp_fig}
\end{figure}

\noindent\textbf{Experimental setup.} To perform real-world data collection, we constructed an experimental prototype as shown in Fig. \ref{fig:exp_fig}. 
Two galvo mirrors raster scan the source and detector separately, emulating measurements that could be obtained with an array of light sources and detectors. A NKT Photonics SuperK EXTREME supercontinuum laser produces either 680 \si{\nano\meter} or 480 \si{\nano\meter}, 80 \si{\mega\hertz} light pulses for absorption and fluorescence experiments, respectively. Photon arrival times are detected using an MPD FastGatedSPAD single-pixel detector with a temporal jitter of \textless 50 \si{\pico\second} connected to a PicoQuant HydraHarp 400. An MPD Picosecond Delayer provides a delay to the synchronization signal from the laser to ensure the SPAD's 5 \si{\nano\second} gate encompasses the entire transient from the scene. To avoid pile-up in the SPAD, we generally operate at low-light conditions, around 1 million counts/sec ($\sim1\%$ of the laser repetition rate).

\noindent\textbf{Scattering tissue phantoms.} \label{sec:materials_scat_phantom}
We use a 3D printer (Formlabs Form 3) to synthesize the optical tissue phantoms used in our experiments. 
Our goal is to emulate a skull phantom and we closely mimic the known properties of the human skull including its thickness and mean scattering length.
The scattering slab is 50 \si{\milli\meter} $\times$ 50 \si{\milli\meter} $\times$ 6.5 \si{\milli\meter} with a scattering coefficient $\mu_s=9 mm^{-1}$, corresponding to $\sim$60 mean free paths (MFPs). Mean free paths is equivalent to mean scattering lengths when absorption is negligible.  
For multiplexing experiments, a thickness of 5 \si{\milli\meter} (corresponding to 45 MFPs) is used. Both the thickness and scattering coefficient of this skull phantom were set to be within the accepted range for human skull \cite{Bevilacqua1999, Li2007}. 
We adapt the procedure used by Dempsey \textit{et al.} and synthesize our own resin for optical phantoms \cite{Dempsey2017}. The scattering parameters of the phantom are set by controlling the volume ratio of the 'white' and 'clear' Form resins. 
The scattering coefficient of the phantom can be determined by measuring the temporal broadening of the transients \cite{Bouchard2010}. Additional information regarding the simulation can be found in our supplementary material, section 2.
In Fig. \ref{fig:08b_res_test}, we see that our experimentally measured transient matches with the output of the Monte Carlo simulation. 
The surface curvature of the human skull is something our skull phantom does not emulate, but we do not believe this has a significant effect on the resolution or performance characteristics predicted by our phantoms.
The results shown in Figs. \ref{fig:08b_res_test},  \ref{fig:multiplexing_result_sim}, \ref{fig:09_exp_fluor_recon}, and \ref{fig:10_exp_confocal_recon} use this skull phantom as the scattering layer between the target and the imaging system.

\begin{figure}[t]
\begin{center}
   \includegraphics[width=1\linewidth]{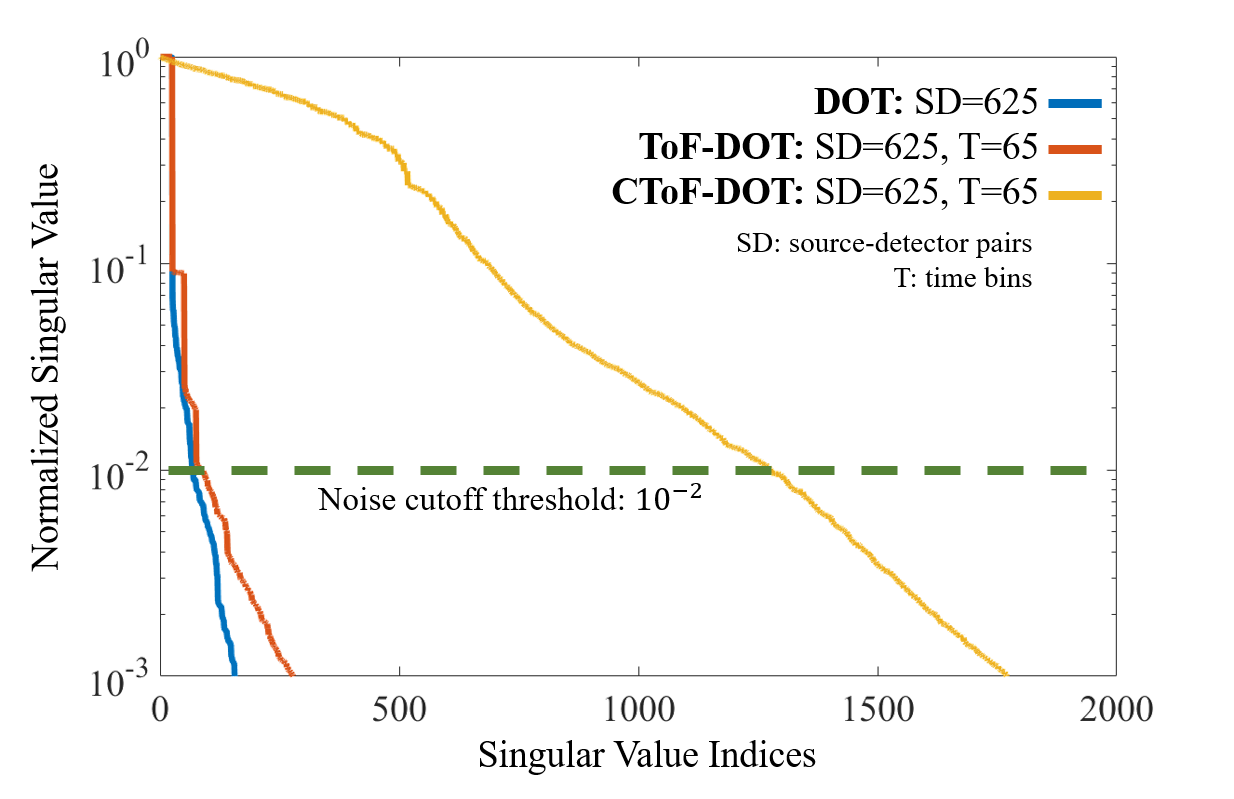}
\end{center}
\caption{\textbf{Jacobian matrix conditioning.} The singular values of the Jacobian matrix are plotted to determine the matrix conditioning. We compare traditional DOT (blue), ToF-DOT (red), and our CToF-DOT (yellow). We see that the introduction of time binning (ToF-DOT) and confocal geometry (CToF-DOT) provides improvements to our matrix conditioning.}
\label{fig:05_singular_vals}
\end{figure}
\noindent\textbf{Absorptive and fluorescent targets.} 
To emulate an absorptive target such as a tumor, we use a E-ink display behind the scattering tissue sample.
An E-ink display is a 2-dimensional monitor, which allows us to programmatically control the spatially varying absorption at a fine spatial resolution. This is achieved by controlling the position of white and black microbeads \cite{Comiskey1998}.
The results shown in Figs. \ref{fig:08b_res_test}, \ref{fig:multiplexing_result_sim}, and \ref{fig:10_exp_confocal_recon} use the E-ink display-based target behind the skull phantom.
In order to emulate fluorescent targets, we embed fluorescent beads (Fluoresbrite YG 1 \si{\micro\meter} beads) in PDMS.
The spatial patterning of the fluorescent target is achieved using a 3D-printed mold. 
The results shown in Fig. \ref{fig:09_exp_fluor_recon} use the fluorescent target behind the skull phantom.

\Section{Results}
We perform an extensive array of experiments, and performance characterizations both in simulation and experimentally using a benchtop prototype ToF-DOT system. 

\begin{figure}[t]
\begin{center}
   \includegraphics[width=1.0\linewidth]{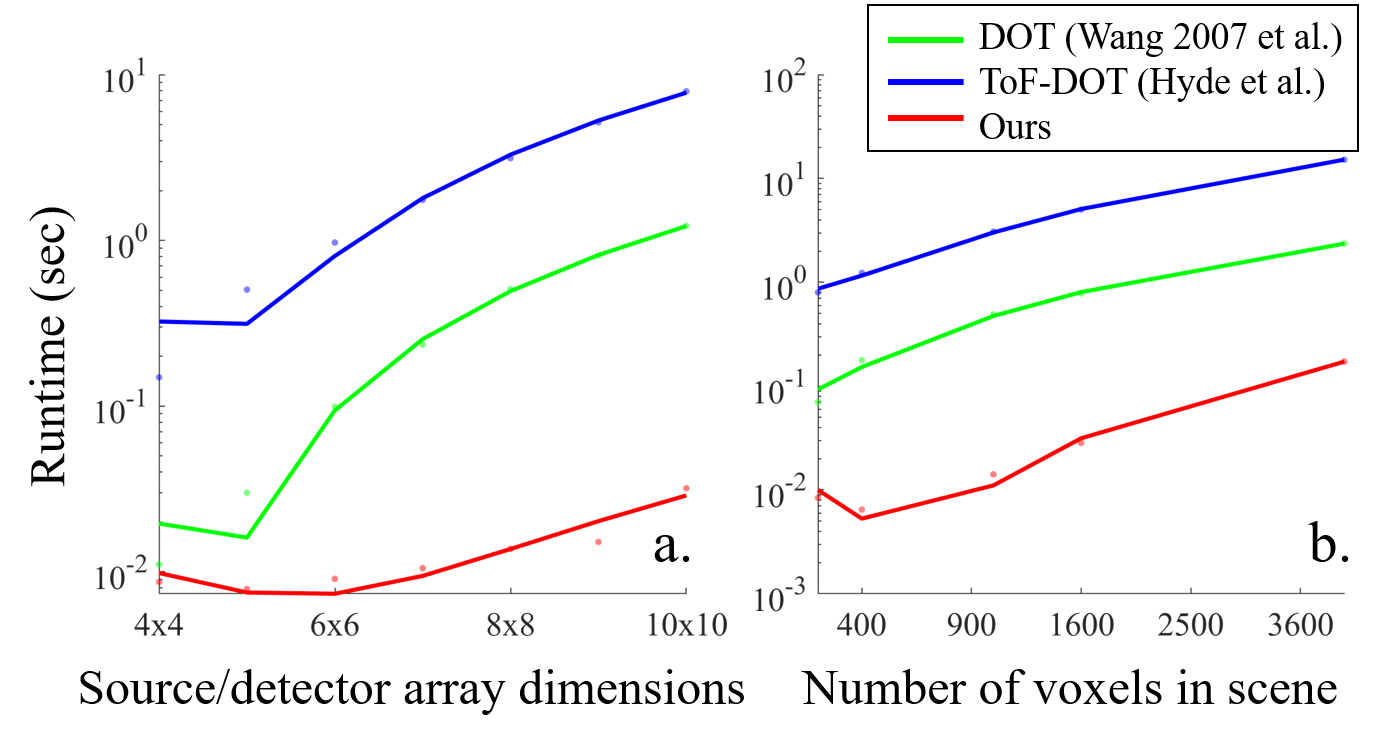}
\end{center}
\caption{\textbf{Algorithm runtime characterization.} The algorithm runtime was characterized as a function of source-detector array size (a), and the voxel grid size (b). We see almost two orders of magnitude decrease in runtime using our methods as compared to traditional DOT \cite{Wang2007} and ToF-DOT \cite{ Hyde2002}.}
\label{fig:06_alg_runtime}
\end{figure}

\SubSection{Conditioning Analysis of CToF-DOT}
\begin{figure}[t]
\begin{center}
   \includegraphics[width=1\linewidth]{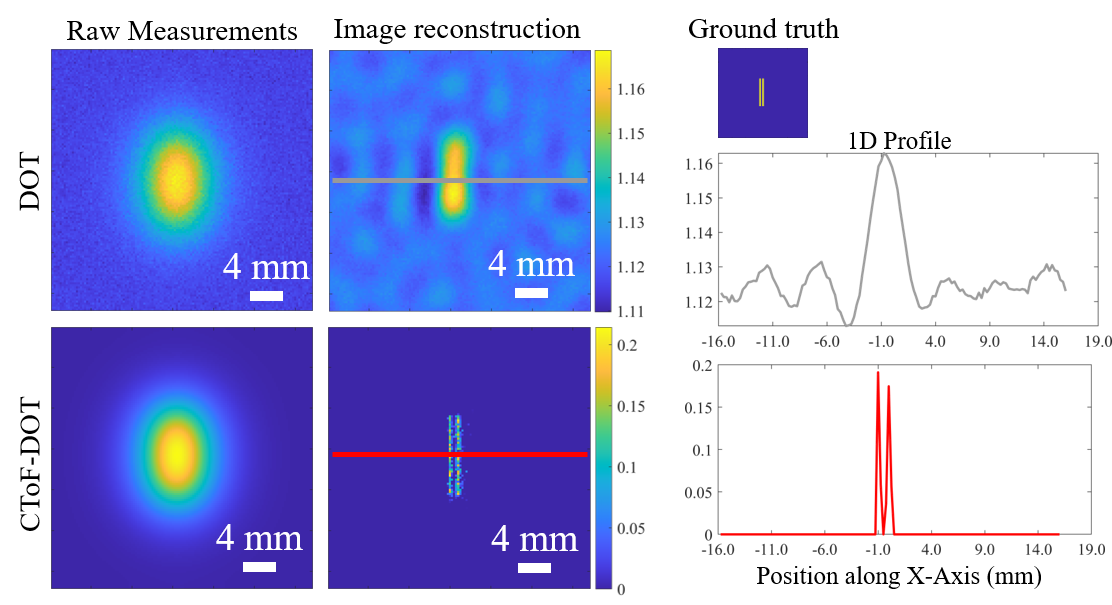}
\end{center}
\caption{\textbf{Simulated spatial resolution of CTOF-DoT.} Our technique is able to resolve two 0.5 mm thick lines separated by 0.5 mm (shown in ground truth image on top right).}
\label{fig:08a_sim_res_test}
\end{figure}

\begin{figure}[t]
\begin{center}
   \includegraphics[width=1\linewidth]{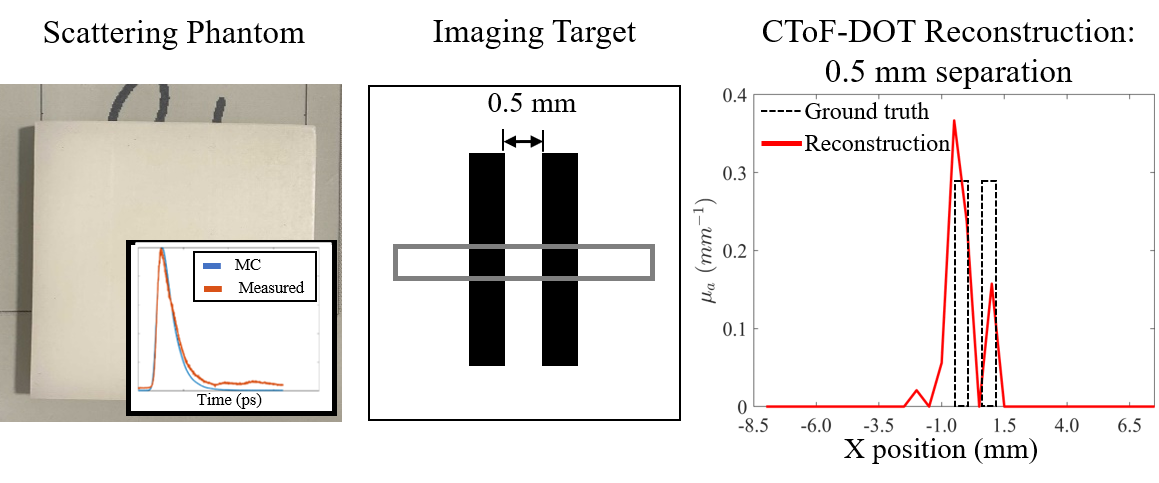}
\end{center}
\caption{\textbf{Resolution test with experimental data.} (Left) Eink target covered by the skull phantom. Inset image shows the matching calibration between Monte Carlo and experimental data, thus verifying the scattering coefficient. (Right) 1D image reconstruction showing our system can resolve two lines separated by 0.5 mm spacing.}
\label{fig:08b_res_test}
\end{figure}

\begin{figure*}[ht]
\centering\includegraphics[width=1.0\linewidth]{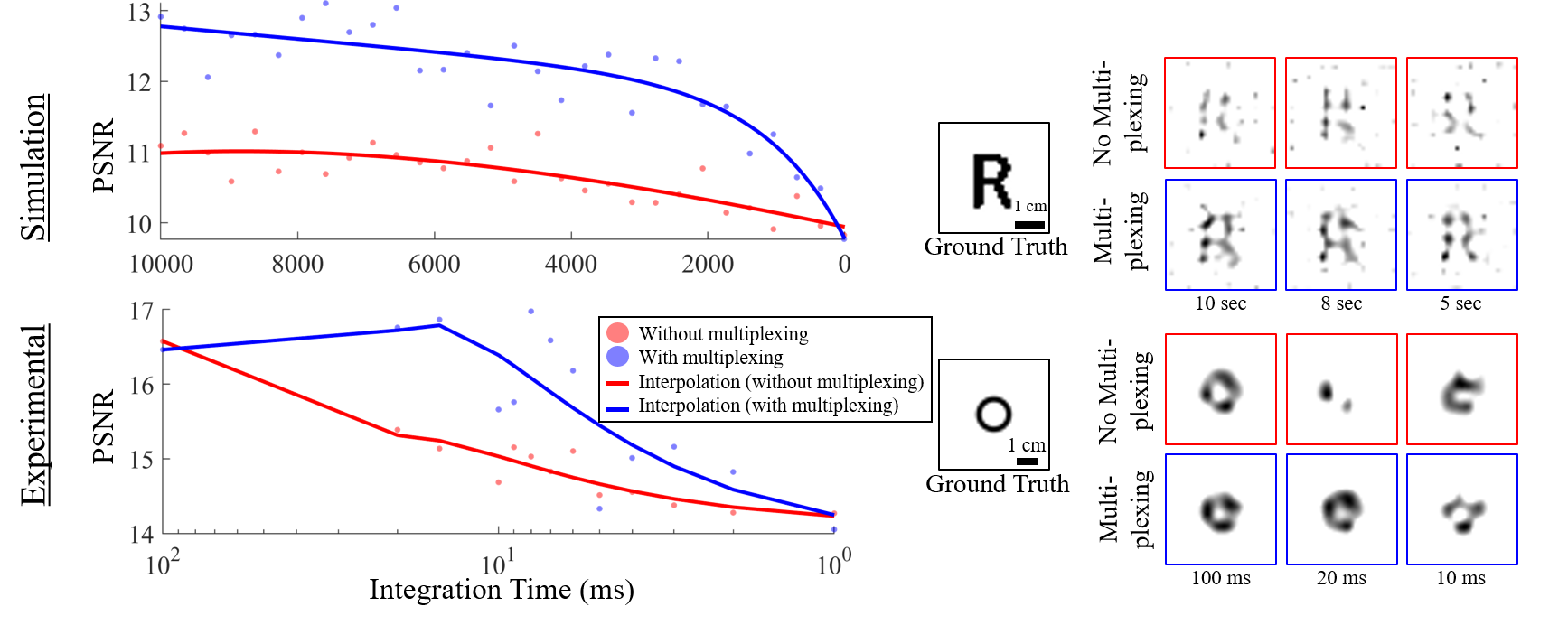}
\caption{\textbf{Simulated and Experimental Multiplexing Results.} Multiplexing allows comparable performance with reduced integration time compared with single point scanning for CToF-DOT imaging. (Left) Plots shows PSNR versus integration time for simulated and experimental results. The images correspond to the image reconstructions performed at different integration times with/without multiplexing. With multiplexing, the image reconstruction is more robust to noise at lower integration times. Measurements were captured through a 5 mm phantom with $\mu_s=9$\si{\milli\meter}$^{-1}$ ($\sim$45 MFPs). } 
\label{fig:multiplexing_result_sim}
\end{figure*}
 Inverting the Jacobian matrix is critical to our image reconstruction procedure. A well-conditioned Jacobian will allow us to improve our image reconstruction quality. As shown in Fig. \ref{fig:05_singular_vals} we demonstrate that the additional information provided by time-binning results in a more well-conditioned matrix. Each Jacobian was obtained through Monte Carlo simulations. We compare three cases: 1) \textit{Traditional DOT} in which all measurements are a scalar intensity value; 2) \textit{ToF-DOT}, which uses all time bins; and 3) \textit{CToF-DOT}. 625 total scan points and 65 time bins were used for each Jacobian. The simulated scene was a $25 \times 25 \times 8$ grid of 1 \si{\milli\meter}$^{3}$ cubes. All singular value plots were normalized to 1. Below a threshold $10^{-2}$, the singular values are considered to be below the noise floor. We see that the introduction of time-domain information improves the matrix conditioning, increasing the minimum singular value from 67 to 82. In addition, for the confocal geometry, because all 625 measurements were collocated, the minimum singular value was further increased to 1276. This provides additional support that collocated source-detector pairs provide more information than an arbitrary set of source-detector pairs. 

\begin{figure}[t]
\begin{center}
   \includegraphics[width=1.0\linewidth]{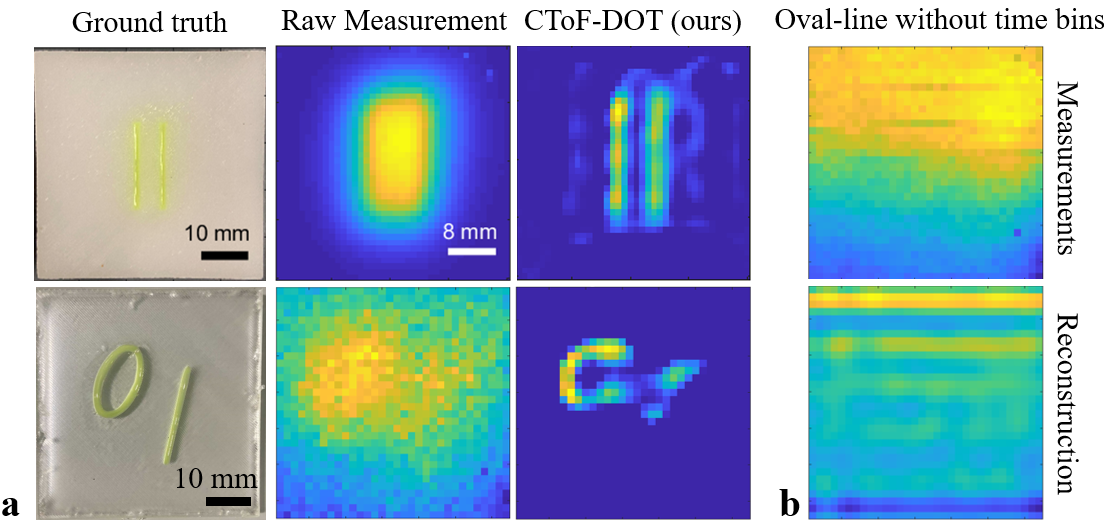}
\end{center}
\caption{\textbf{Fluorescence imaging with CToF-DOT.} (a) Image reconstruction of fluorescence targets (2 4 \si{\milli\meter} lines separated by 4 \si{\milli\meter} and oval-line scene) using CToF-DOT. (b) For the oval-line image reconstruction, we show that the image cannot be recovered without time binning.}
\label{fig:09_exp_fluor_recon}
\end{figure}

\begin{figure}[t]
\centering\includegraphics[width=1.0\linewidth]{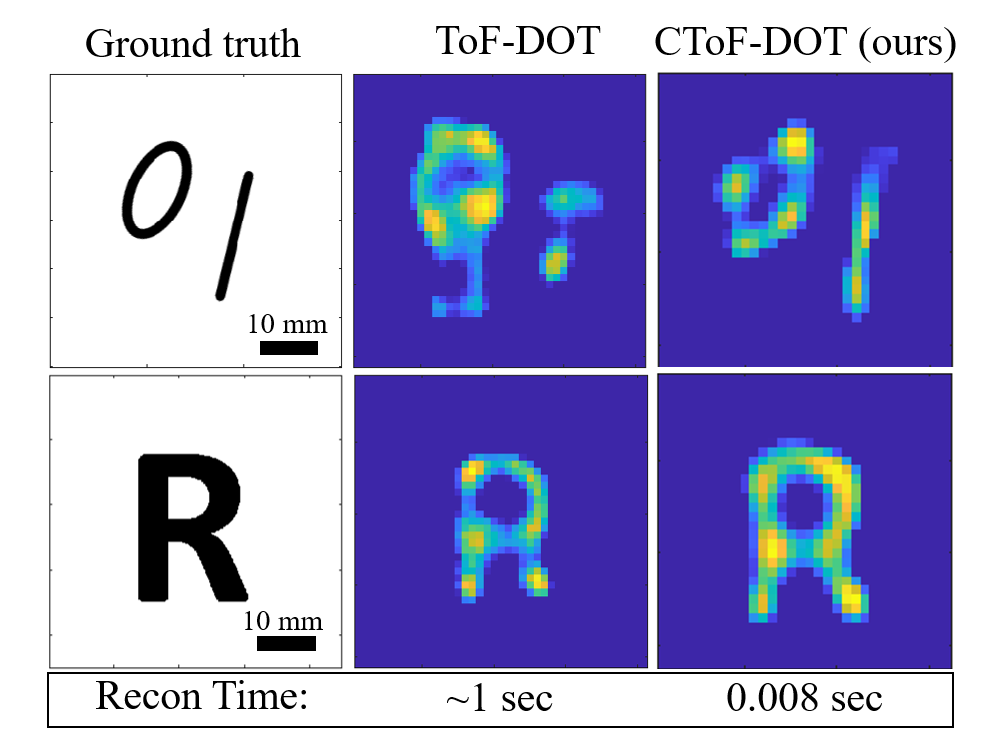}
\caption{\textbf{Experimental image reconstruction of absorptive targets:} Comparison of ToF-DOT and CToF-DOT reconstruction of 2D absorption targets through a 6.5 mm phantom with $\mu_s=9$\si{\milli\meter}$^{-1}$ ($\sim$60 MFPs). We maintain comparable image reconstruction quality while reducing the computation time for the inverse solver by approximately two orders of magnitude, and reducing the number of scan points by almost an order of magnitude.
}
\label{fig:10_exp_confocal_recon}
\end{figure}

\begin{figure*}[t]
\begin{center}
   \includegraphics[width=1.0\linewidth]{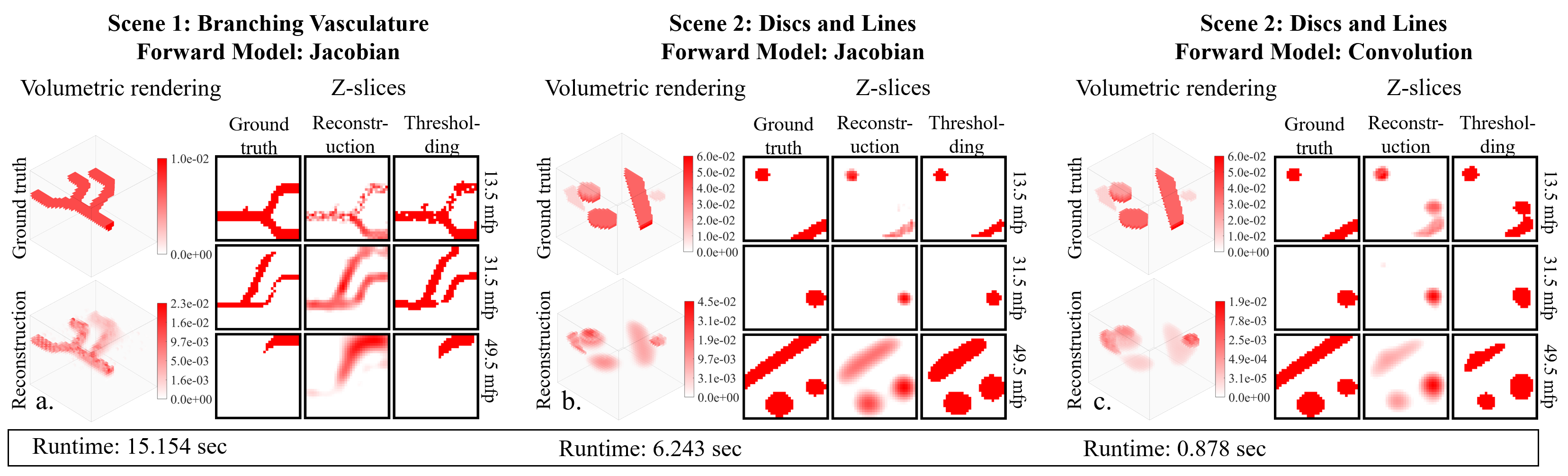}
\end{center}
\caption{\textbf{Simulated 3D Image Reconstruction:} 3D Image reconstruction on simulated data. We show both the volumetric rendering as well as z-slices at multiple depths. The corresponding optical depth for each z-slice is listed to the right of each image. For each z-slice, we show the ground truth image, the image reconstruction using FISTA, and the reconstructed image with a layer-wise threshold.}
\label{fig:13_3D_sim_im_recon}
\end{figure*}

\SubSection{Reconstruction Speed}

Additionally, we test the algorithm runtime. These experiments were conducted on an Intel Xeon 3.30 GHz CPU. We test how the image reconstruction speed is affected by 2 system parameters: the voxel size (for a fixed total area), and the number of sources and detectors. 
We compared CToF-DOT to the algorithms for traditional DOT and ToF-DOT, which were constructed in-house (described by Wang \textit{et al.} and Hyde \textit{et al.}, respectively \cite{Wang2007, Hyde2002}).
In Fig. \ref{fig:06_alg_runtime}, we see that the confocal geometry achieves almost 2 orders of magnitude improvements in speed primarily attributed to a reduction in Jacobian matrix size. Here, we see that CToF-DOT is even faster than traditional DOT. This is because the increase in runtime due to using additional time bins in CToF-DOT is outweighed by the speed-ups from considering only collocated source-detector pairs.

\SubSection{Spatial Resolution Tests}

\noindent\textbf{Simulation resolution test.} We test the spatial resolution that can be achieved by traditional DOT and our method (Fig. \ref{fig:08a_sim_res_test}). The simulated scene consists of two fluorescent lines, with 0.5 mm line width and separation. The target scene corresponds to a 2-dimensional voxel grid of $128\times128$ voxels spanning a total volume of $32\times32\si{\milli\meter}$, embedded 6.5 mm deep in a homogeneous scattering media with a background scattering coefficient set to $\mu_s=9.0$\si{\milli\meter}$^{-1}$. With a $128\times128$ confocal scan and 0 \si{\milli\meter} separation between the source and detector, we see that we are able to clearly resolve the two lines, which indicates our system can resolve \si{\milli\meter}-scale features. For comparison, we simulated a set of DOT measurements with 4\si{\milli\meter} separation between the source and detector, still using only adjacent source-detector pairs, and no temporal data, a total of $128^2$ values in the measurement. We see that time-domain data allows us to achieve higher resolution reconstructions. This is primarily because, by using time-gating, CToF-DOT is more effective in removing excitation light, which primarily acts as a source of noise. For both DOT and CToF-DOT, the total measurements, which are a sum of the excitation and emission light are normalized to the same value, based on the exposure duration. By using time-gating, CToF-DOT rejects more of the excitation light, while passing the emission light, which is stronger at later time bins due to the fluorescence lifetime. 

\noindent\textbf{Experimental resolution test.}
In Fig. \ref{fig:08b_res_test}, we performed a 1-dimensional resolution test through a 6.5 \si{\milli\meter} thick skull phantom ($\mu_s=9$\si{\milli\meter}$^{-1}$) by scanning 32 points in a confocal geometry. To obtain the Jacobian experimentally, a black line was projected on the E-ink display at 32 locations with 0.5 \si{\milli\meter} separation. For each line position, a 32-point scan was captured, which becomes a column of the Jacobian matrix. After obtaining the Jacobian, we projected the target image onto the E-ink display: two lines of thickness and separation distance 0.5 \si{\milli\meter}. Though there is a slight offset due to calibration, we are able to resolve the two lines and demonstrate \si{\milli\meter}-scale spatial resolution (Fig. \ref{fig:08b_res_test}). Additionally, in our supplementary material, we show the results for resolving 0.5 \si{\milli\meter}, 1.0 \si{\milli\meter}, 2.0 \si{\milli\meter} thick lines.

\SubSection{Advantages of Multiplexing}

\noindent\textbf{Simulations on multiplexing.}
We tested source multiplexing with an $8\times 8$ array of sources, which leads to multiplexing with a $64 \times 64$ Hadamard matrix. 
The simulated measurements and Jacobian were generated using the analytical expressions. 
Poisson noise was applied, assuming a count rate of 5 million counts per second, the approximate intensity level before our SPAD experiences the pile-up effect. 
In addition, dark count noise was added to our measurements, with a rate parameter of 200 counts/sec corresponding to the dark count rate of the FastGatedSPAD. 
Image reconstruction of the letter 'R' is performed for a range of integration times. 
In Fig. \ref{fig:multiplexing_result_sim}, we see that with multiplexing, the image reconstruction still maintains a reasonable PSNR at short exposure durations. 
This demonstrates that the image reconstruction with multiplexing is more robust to increased noise at lower integration times. 
The panels on the right of Fig. \ref{fig:multiplexing_result_sim} show the reconstruction results, again showing increased robustness to noise with multiplexing.

\noindent\textbf{Multiplexing with experimental data.}
In addition to simulations, we also captured experimental data to test the benefits of source multiplexing. The plots on Fig. \ref{fig:multiplexing_result_sim} show the PSNR as a function of the integration time. We show that the multiplexed measurements are more robust to higher noise levels at lower integration times. Therefore, multiplexing can improve the temporal resolution by reducing the integration time needed to maintain a threshold image reconstruction quality. From the experimental results, we see an order of magnitude improvement since the image reconstruction quality at 10 \si{\milli\second} is approximately comparable to 100 ms integration time without multiplexing (Fig. \ref{fig:multiplexing_result_sim}). We must also account for a factor of $\frac{1}{2}$ since multiplexing doubles the total number of measurements, resulting in an overall $\sim$5 times reduction in measurement capture time. 

\SubSection{Additional Image Reconstructions}
\label{sec:image_recon_exp}
In the image reconstruction experiments with real-world data, the scattering slab is placed on top of either the E-ink display or the fluorescent target to emulate scattering by biological tissue.
In Fig. \ref{fig:09_exp_fluor_recon}(a). we see that the signal from the fluorescent targets is significantly enhanced by removing early-arriving photons using time-gating, albeit still blurred due to the effects of scattering. The target images can be recovered by our image reconstruction algorithm.
Fig. \ref{fig:09_exp_fluor_recon}(b) shows measurements and a reconstruction of the oval-line scene without time-binning. The measurements are dominated by background noise from the excitation light leading to an image reconstruction that contains virtually no information about the underlying scene. 
The oval line scene experiment was conducted with the standard 6.5 mm phantom with $\mu_s=9$\si{\milli\meter}$^{-1}$, without an emission filter. However, for imaging the two lines image reconstruction, a filter was used as well as an un-calibrated skull phantom with scattering coefficient in the range $5-10$\si{\milli\meter}$^{-1}$.

\begin{figure}[t]
\begin{center}
   \includegraphics[width=1\linewidth]{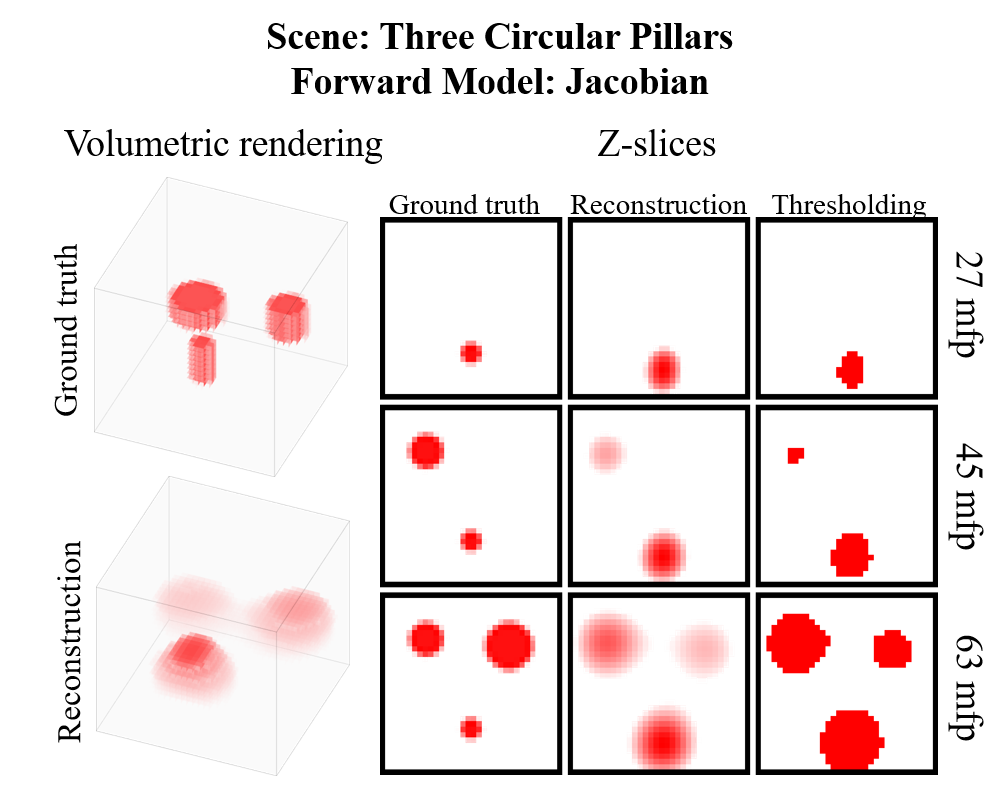}
\end{center}
\caption{\textbf{3D Image Reconstruction on Experimental Data} Our 3D image reconstruction is applied to experimental data. The ground truth imaging target consists of three pillars with varying depths and diameters.}
\label{fig:14_3D_exp_im_recon}
\end{figure}

We show the benefits of using a confocal geometry in Fig. \ref{fig:10_exp_confocal_recon}(a). Here, we image absorber targets displayed with the E-ink display. In the standard TD-DOT setup, we capture a measurement for all pairs of sources and detectors. For this experiment, we use a $10\times10$ array of sources and detectors, which correspond to 10,000 total scan points. Additionally, we capture measurements for the same scene using a $32\times32$ array in the confocal geometry, corresponding to 1024 scan points. As shown in Fig. \ref{fig:10_exp_confocal_recon}(a), even though the number of scan points is reduced by almost an order of magnitude, we are able to maintain comparable image reconstruction quality using confocal ToF-DOT compared to ToF-DOT. Additionally, with the confocal geometry, the algorithm runtime is reduced by approximately two orders of magnitude from $\sim$1 sec to $\sim$8 ms. 

\noindent\textbf{3D Image Reconstruction} In addition to 2D image reconstruction with an Eink display, we also performed 3D image reconstruction on simulated and real-world data. Each scene occupies a total area of $32\text{mm}\times32\text{mm}$. In fig. \ref{fig:13_3D_sim_im_recon}a., the imaging target is an artificial vasculature at multiple depths. The optical absorption is assumed to be homogeneous for this scene. In fig. \ref{fig:13_3D_sim_im_recon}b., c. the imaging target is a set of discs and lines at multiple depths, with absorption coefficient increasing by $0.01\si{\milli\meter}^{-1}$ per layer from $\mu_a=0.01\si{\milli\meter}^{-1}$ at the shallowest layer to $\mu_a=0.06\si{\milli\meter}^{-1}$. We also compare the Jacobian forward model (panel b.) with the convolutional forward model (panel c). Both algorithms used the same number of iterations of FISTA and achieve comparable resulting image quality; however, the convolutional model performs almost an order of magnitude faster.

Our experimental scene consists of a 3D printed black mold containing circular pillars of diameters 6 \si{\milli\meter}, 10 \si{\milli\meter}, and 14\si{\milli\meter}, embedded at depths of 2 \si{\milli\meter}, 4\si{\milli\meter}, and 6\si{\milli\meter}, respectively (fig. \ref{fig:14_3D_exp_im_recon}). A liquid skull phantom, with the same optical properties as described in section \ref{sec:materials_scat_phantom}, was poured over these features to emulate biological features embedded in a scattering background. To capture the measurements, we scanned 32$\times$32 collocated source-detector pairs uniformly distributed over an area of $40 \si{\milli\meter}\times 40\si{\milli\meter}$. For reconstructing experimental data, we used a timing resolution of 200\si{\pico\second}. As shown in fig. \ref{fig:14_3D_exp_im_recon}, using our CToF-DOT algorithm, we are able to distinguish the three cylinders and correctly localize their depth. 

\Section{Conclusions}
We demonstrate that confocal and multiplexed versions of ToF-DOT have the potential to achieve millimeter resolution, real-time imaging through thick scattering tissue. 
With future developments in terms of on-chip SPAD hardware and integrated source-detector arrays, these results can lead to wearable imaging devices paving the way for high resolution structural and functional imaging of the brain. 

\ifpeerreview \else
\Section{Acknowledgments}
The authors would like to thank Prof. Aswin Sankaranarayanan for his suggestions and edits to our paper. The authors would like to acknowledge Biorender for use of its icons to help in figure design. This research was partially funded by the Defense Advanced Research Projects Agency (DARPA), Contract No. N66001-19-C-4020 and by the NSF Expeditions in Computing Grant \#1730147; in addition, Y. Zhao was supported by a training fellowship from the NLM Training Program (T15LM007093). The views, opinions and/or findings expressed are those of the author and should not be interpreted as representing the official views or policies of the Department of Defense or the U.S. Government.
\fi

\bibliographystyle{IEEEtran}
\bibliography{dotbib}

\begin{thebibliography}{10}
\providecommand{\url}[1]{#1}
\csname url@samestyle\endcsname
\providecommand{\newblock}{\relax}
\providecommand{\bibinfo}[2]{#2}
\providecommand{\BIBentrySTDinterwordspacing}{\spaceskip=0pt\relax}
\providecommand{\BIBentryALTinterwordstretchfactor}{4}
\providecommand{\BIBentryALTinterwordspacing}{\spaceskip=\fontdimen2\font plus
\BIBentryALTinterwordstretchfactor\fontdimen3\font minus
  \fontdimen4\font\relax}
\providecommand{\BIBforeignlanguage}[2]{{%
\expandafter\ifx\csname l@#1\endcsname\relax
\typeout{** WARNING: IEEEtran.bst: No hyphenation pattern has been}%
\typeout{** loaded for the language `#1'. Using the pattern for}%
\typeout{** the default language instead.}%
\else
\language=\csname l@#1\endcsname
\fi
#2}}
\providecommand{\BIBdecl}{\relax}
\BIBdecl

\bibitem{Pediredla2016}
A.~K. Pediredla, S.~Zhang, B.~Avants, F.~Ye, S.~Nagayama, Z.~Chen, C.~Kemere,
  J.~T. Robinson, and A.~Veeraraghavan, ``{Deep imaging in scattering media
  with selective plane illumination microscopy},'' \emph{Journal of Biomedical
  Optics}, vol.~21, no.~12, pp. 1--14, 2016.

\bibitem{oh2019skin}
B.-H. Oh, K.~H. Kim, and K.-Y. Chung, ``Skin imaging using ultrasound imaging,
  optical coherence tomography, confocal microscopy, and two-photon microscopy
  in cutaneous oncology,'' \emph{Frontiers in Medicine}, vol.~6, p. 274, 2019.

\bibitem{Bevilacqua1999}
F.~Bevilacqua, D.~Piguet, P.~Marquet, J.~D. Gross, B.~J. Tromberg, and
  C.~Depeursinge, ``{In vivo local determination of tissue optical properties:
  applications to human brain},'' \emph{Applied Optics}, vol.~38, no.~22, pp.
  4939--4950, 1999.

\bibitem{boas2001imaging}
D.~A. Boas, D.~H. Brooks, E.~L. Miller, C.~A. DiMarzio, M.~Kilmer, R.~J.
  Gaudette, and Q.~Zhang, ``Imaging the body with diffuse optical tomography,''
  \emph{IEEE signal processing magazine}, vol.~18, no.~6, pp. 57--75, 2001.

\bibitem{Xia2014}
J.~Xia, J.~Yao, and L.~V. Wang, ``\BIBforeignlanguage{eng}{{Photoacoustic
  tomography: principles and advances}},''
  \emph{\BIBforeignlanguage{eng}{Electromagnetic waves (Cambridge, Mass.)}},
  vol. 147, pp. 1--22, 2014.

\bibitem{puszka2013time}
A.~Puszka, L.~Di~Sieno, A.~Dalla~Mora, A.~Pifferi, D.~Contini, G.~Boso,
  A.~Tosi, L.~Herv{\'e}, A.~Planat-Chr{\'e}tien, A.~Koenig \emph{et~al.},
  ``Time-resolved diffuse optical tomography using fast-gated single-photon
  avalanche diodes,'' \emph{Biomedical optics express}, vol.~4, no.~8, pp.
  1351--1365, 2013.

\bibitem{Nie2012}
L.~Nie, X.~Cai, K.~Maslov, A.~Garcia-Uribe, M.~A. Anastasio, and L.~V. Wang,
  ``\BIBforeignlanguage{eng}{{Photoacoustic tomography through a whole adult
  human skull with a photon recycler}},''
  \emph{\BIBforeignlanguage{eng}{Journal of biomedical optics}}, vol.~17,
  no.~11, p. 110506, nov 2012.

\bibitem{Wang2007}
L.~Wang and H.-I. Wu, \emph{{Biomedical Optics: Principles and Imaging}}.\hskip
  1em plus 0.5em minus 0.4em\relax Hoboken, NJ: John Wiley and Sons, 2007.

\bibitem{kempe1996comparative}
M.~Kempe, W.~Rudolph, and E.~Welsch, ``Comparative study of confocal and
  heterodyne microscopy for imaging through scattering media,'' \emph{JOSA A},
  vol.~13, no.~1, pp. 46--52, 1996.

\bibitem{sergeeva2010scattering}
E.~A. Sergeeva, ``Scattering effect on the imaging depth limitin two-photon
  fluorescence microscopy,'' \emph{Quantum Electronics}, vol.~40, no.~5, p.
  411, 2010.

\bibitem{Liu2020}
C.~Liu, A.~Maity, A.~W. Dubrawski, A.~Sabharwal, and S.~G. Narasimhan, ``High
  resolution diffuse optical tomography using short range indirect subsurface
  imaging,'' in \emph{IEEE International Conference on Computational
  Photography}.\hskip 1em plus 0.5em minus 0.4em\relax IEEE, May 2020.

\bibitem{zhao2017review}
H.~Zhao and R.~J. Cooper, ``Review of recent progress toward a fiberless,
  whole-scalp diffuse optical tomography system,'' \emph{Neurophotonics},
  vol.~5, no.~1, p. 011012, 2017.

\bibitem{frijia2020functional}
E.~M. Frijia, A.~Billing, S.~Lloyd-Fox, E.~V. Rosas, L.~Collins-Jones, M.~M.
  Crespo-Llado, M.~P. Amad{\'o}, T.~Austin, A.~Edwards, L.~Dunne \emph{et~al.},
  ``Functional imaging of the developing brain with wearable high-density
  diffuse optical tomography: a new benchmark for infant neuroimaging outside
  the scanner environment,'' \emph{NeuroImage}, p. 117490, 2020.

\bibitem{wheelock2019high}
M.~D. Wheelock, J.~P. Culver, and A.~T. Eggebrecht, ``High-density diffuse
  optical tomography for imaging human brain function,'' \emph{Review of
  Scientific Instruments}, vol.~90, no.~5, p. 051101, 2019.

\bibitem{gibson2005recent}
A.~Gibson, J.~Hebden, and S.~R. Arridge, ``Recent advances in diffuse optical
  imaging,'' \emph{Physics in Medicine \& Biology}, vol.~50, no.~4, p.~R1,
  2005.

\bibitem{pifferi2016new}
A.~Pifferi, D.~Contini, A.~Dalla~Mora, A.~Farina, L.~Spinelli, and
  A.~Torricelli, ``New frontiers in time-domain diffuse optics, a review,''
  \emph{Journal of biomedical optics}, vol.~21, no.~9, p. 091310, 2016.

\bibitem{Lyons2019}
A.~Lyons, F.~Tonolini, A.~Boccolini, A.~Repetti, R.~Henderson, Y.~Wiaux, and
  D.~Faccio, ``{Computational time-of-flight diffuse optical tomography},''
  \emph{Nature Photonics}, vol.~13, no.~8, pp. 575--579, 2019.

\bibitem{farina2017time}
A.~Farina, S.~Tagliabue, L.~Di~Sieno, E.~Martinenghi, T.~Durduran, S.~Arridge,
  F.~Martelli, A.~Torricelli, A.~Pifferi, and A.~Dalla~Mora, ``Time-domain
  functional diffuse optical tomography system based on fiber-free silicon
  photomultipliers,'' \emph{Applied Sciences}, vol.~7, no.~12, p. 1235, 2017.

\bibitem{di2017miniaturized}
L.~Di~Sieno, J.~Nissinen, L.~Hallman, E.~Martinenghi, D.~Contini, A.~Pifferi,
  J.~Kostamovaara, and A.~D. Mora, ``Miniaturized pulsed laser source for
  time-domain diffuse optics routes to wearable devices,'' \emph{Journal of
  biomedical optics}, vol.~22, no.~8, p. 085004, 2017.

\bibitem{Lindell2020}
D.~B. Lindell and G.~Wetzstein, ``{Three-dimensional imaging through scattering
  media based on confocal diffuse tomography},'' \emph{Nature Communications},
  vol.~11, no.~1, p. 4517, 2020.

\bibitem{puszka2015spatial}
A.~Puszka, L.~Di~Sieno, A.~Dalla~Mora, A.~Pifferi, D.~Contini,
  A.~Planat-Chr{\'e}tien, A.~Koenig, G.~Boso, A.~Tosi, L.~Herv{\'e}
  \emph{et~al.}, ``Spatial resolution in depth for time-resolved diffuse
  optical tomography using short source-detector separations,''
  \emph{Biomedical optics express}, vol.~6, no.~1, pp. 1--10, 2015.

\bibitem{Kim2008}
\BIBentryALTinterwordspacing
H.~K. Kim and A.~H. Hielscher, ``{A PDE-constrained SQP algorithm for optical
  tomography based on the frequency-domain equation of radiative transfer},''
  \emph{Inverse Problems}, vol.~25, no.~1, p. 15010, 2008. [Online]. Available:
  \url{http://dx.doi.org/10.1088/0266-5611/25/1/015010}
\BIBentrySTDinterwordspacing

\bibitem{boas2001simultaneous}
D.~A. Boas, T.~Gaudette, and S.~R. Arridge, ``Simultaneous imaging and optode
  calibration with diffuse optical tomography,'' \emph{Optics express}, vol.~8,
  no.~5, pp. 263--270, 2001.

\bibitem{tarvainen2010corrections}
T.~Tarvainen, V.~Kolehmainen, J.~P. Kaipio, and S.~R. Arridge, ``Corrections to
  linear methods for diffuse optical tomography using approximation error
  modelling,'' \emph{Biomedical Optics Express}, vol.~1, no.~1, pp. 209--222,
  2010.

\bibitem{naser2015time}
M.~A. Naser and M.~J. Deen, ``Time-domain diffuse optical tomography using
  recursive direct method of calculating jacobian at selected temporal
  points,'' \emph{Biomedical Physics \& Engineering Express}, vol.~1, no.~4, p.
  045207, 2015.

\bibitem{mozumder2020time}
M.~Mozumder and T.~Tarvainen, ``Time-domain diffuse optical tomography
  utilizing truncated fourier series approximation,'' \emph{JOSA A}, vol.~37,
  no.~2, pp. 182--191, 2020.

\bibitem{gkioulekas2013inverse}
I.~Gkioulekas, S.~Zhao, K.~Bala, T.~Zickler, and A.~Levin, ``Inverse volume
  rendering with material dictionaries,'' \emph{ACM Transactions on Graphics
  (TOG)}, vol.~32, no.~6, pp. 1--13, 2013.

\bibitem{gkioulekas2016evaluation}
I.~Gkioulekas, A.~Levin, and T.~Zickler, ``An evaluation of computational
  imaging techniques for heterogeneous inverse scattering,'' in \emph{European
  Conference on Computer Vision}.\hskip 1em plus 0.5em minus 0.4em\relax
  Springer, 2016, pp. 685--701.

\bibitem{nimier2019mitsuba}
M.~Nimier-David, D.~Vicini, T.~Zeltner, and W.~Jakob, ``Mitsuba 2: A
  retargetable forward and inverse renderer,'' \emph{ACM Transactions on
  Graphics (TOG)}, vol.~38, no.~6, pp. 1--17, 2019.

\bibitem{nimier2020radiative}
M.~Nimier-David, S.~Speierer, B.~Ruiz, and W.~Jakob, ``Radiative
  backpropagation: an adjoint method for lightning-fast differentiable
  rendering,'' \emph{ACM Transactions on Graphics (TOG)}, vol.~39, no.~4, pp.
  146--1, 2020.

\bibitem{azinovic2019inverse}
D.~Azinovic, T.-M. Li, A.~Kaplanyan, and M.~Nie{\ss}ner, ``Inverse path tracing
  for joint material and lighting estimation,'' in \emph{Proceedings of the
  IEEE/CVF Conference on Computer Vision and Pattern Recognition}, 2019, pp.
  2447--2456.

\bibitem{che2018inverse}
C.~Che, F.~Luan, S.~Zhao, K.~Bala, and I.~Gkioulekas, ``Inverse transport
  networks,'' \emph{arXiv preprint arXiv:1809.10820}, 2018.

\bibitem{levis2015airborne}
A.~Levis, Y.~Y. Schechner, A.~Aides, and A.~B. Davis, ``Airborne
  three-dimensional cloud tomography,'' in \emph{Proceedings of the IEEE
  International Conference on Computer Vision}, 2015, pp. 3379--3387.

\bibitem{levis2017multiple}
A.~Levis, Y.~Y. Schechner, and A.~B. Davis, ``Multiple-scattering microphysics
  tomography,'' in \emph{Proceedings of the IEEE Conference on Computer Vision
  and Pattern Recognition}, 2017, pp. 6740--6749.

\bibitem{satat2018towards}
G.~Satat, M.~Tancik, and R.~Raskar, ``Towards photography through realistic
  fog,'' in \emph{2018 IEEE International Conference on Computational
  Photography (ICCP)}.\hskip 1em plus 0.5em minus 0.4em\relax IEEE, 2018, pp.
  1--10.

\bibitem{Kak1988}
A.~Kak and M.~Slaney, ``Principles of computerized tomographic imaging ieee
  press,'' \emph{New York}, 1988.

\bibitem{Yao2018}
R.~Yao, X.~Intes, and Q.~Fang, ``{Direct approach to compute Jacobians for
  diffuse optical tomography using perturbation Monte Carlo-based photon
  “replay”},'' \emph{Biomedical Optics Express}, vol.~9, no.~10, pp.
  4588--4603, 2018.

\bibitem{Hyde2002}
D.~Hyde, ``Improving forward matrix generation and utilization for time domain
  diffuse optical tomography,'' Ph.D. dissertation, Worcester Polytechnic
  Institute, 2002.

\bibitem{Kim2017}
H.~K. Kim, L.~D. Montejo, J.~Jia, and A.~H. Hielscher,
  ``\BIBforeignlanguage{eng}{{Frequency-domain optical tomographic image
  reconstruction algorithm with the simplified spherical harmonics (SP(3))
  light propagation model.}}'' \emph{\BIBforeignlanguage{eng}{Int J Therm
  Sci.}}, vol. 116, pp. 265--277, jun 2017.

\bibitem{Wang1995}
L.~{Wang, L., Jacques, S. L., {\&} Zheng}, ``{MCML - Monte Carlo modeling of
  light transport in multi-layered tissues},'' \emph{Computer methods and
  programs in biomedicine}, vol.~47, no.~2, pp. 131--146, 1995.

\bibitem{Beck2009}
A.~Beck and M.~Teboulle, ``{A Fast Iterative Shrinkage-Thresholding
  Algorithm},'' \emph{Society for Industrial and Applied Mathematics Journal on
  Imaging Sciences}, vol.~2, no.~1, pp. 183--202, 2009.

\bibitem{Velten2012}
A.~Velten, T.~Willwacher, O.~Gupta, A.~Veeraraghavan, M.~G. Bawendi, and
  R.~Raskar, ``{Recovering three-dimensional shape around a corner using
  ultrafast time-of-flight imaging},'' \emph{Nature Communications}, vol.~3,
  no.~1, p. 745, 2012.

\bibitem{Pediredla2019}
A.~{Pediredla}, A.~{Dave}, and A.~{Veeraraghavan}, ``Snlos: Non-line-of-sight
  scanning through temporal focusing,'' in \emph{2019 IEEE International
  Conference on Computational Photography (ICCP)}, 2019, pp. 1--13.

\bibitem{Ahn2019}
B.~Ahn, A.~Dave, A.~Veeraraghavan, I.~Gkioulekas, and A.~Sankaranarayanan,
  ``{Convolutional Approximations to the General Non-Line-of-Sight Imaging
  Operator},'' in \emph{2019 IEEE/CVF International Conference on Computer
  Vision (ICCV)}, 2019, pp. 7888--7898.

\bibitem{OToole2018}
M.~O'Toole, D.~B. Lindell, and G.~Wetzstein, ``{Confocal non-line-of-sight
  imaging based on the light-cone transform},'' \emph{Nature}, vol. 555, no.
  7696, pp. 338--341, 2018.

\bibitem{pifferi2008time}
A.~Pifferi, A.~Torricelli, L.~Spinelli, D.~Contini, R.~Cubeddu, F.~Martelli,
  G.~Zaccanti, A.~Tosi, A.~Dalla~Mora, F.~Zappa \emph{et~al.}, ``Time-resolved
  diffuse reflectance using small source-detector separation and fast
  single-photon gating,'' \emph{Physical review letters}, vol. 100, no.~13, p.
  138101, 2008.

\bibitem{torricelli2005time}
A.~Torricelli, A.~Pifferi, L.~Spinelli, R.~Cubeddu, F.~Martelli, S.~Del~Bianco,
  and G.~Zaccanti, ``Time-resolved reflectance at null source-detector
  separation: improving contrast and resolution in diffuse optical imaging,''
  \emph{Physical review letters}, vol.~95, no.~7, p. 078101, 2005.

\bibitem{Cossairt2013}
O.~{Cossairt}, M.~{Gupta}, and S.~K. {Nayar}, ``When does computational imaging
  improve performance?'' \emph{IEEE Transactions on Image Processing}, vol.~22,
  no.~2, pp. 447--458, 2013.

\bibitem{sankaranarayanan2018hadamard}
A.~C. Sankaranarayanan, ``Hadamard multiplexing and when it is useful,'' 2018.

\bibitem{Schechner2007}
Y.~Y. Schechner, S.~K. Nayar, and P.~N. Belhumeur, ``{Multiplexing for Optimal
  Lighting},'' \emph{IEEE Transactions on Pattern Analysis and Machine
  Intelligence}, vol.~29, no.~8, pp. 1339--1354, aug 2007.

\bibitem{Mitra2014}
K.~{Mitra}, O.~S. {Cossairt}, and A.~{Veeraraghavan}, ``A framework for
  analysis of computational imaging systems: Role of signal prior, sensor noise
  and multiplexing,'' \emph{IEEE Transactions on Pattern Analysis and Machine
  Intelligence}, vol.~36, no.~10, pp. 1909--1921, 2014.

\bibitem{Liebert2008}
A.~Liebert, H.~Wabnitz, N.~{\.{Z}}o{\l}ek, and R.~Macdonald, ``{Monte Carlo
  algorithm for efficient simulation of time-resolved fluorescence in layered
  turbid media},'' \emph{Optics Express}, vol.~16, no.~17, pp.
  13\,188--13\,202, 2008.

\bibitem{Chen2011}
J.~Chen, V.~Venugopal, and X.~Intes, ``{Monte Carlo based method for
  fluorescence tomographic imaging with lifetime multiplexing using time
  gates},'' \emph{Biomedical Optics Express}, vol.~2, no.~4, pp. 871--886,
  2011.

\bibitem{Li2007}
H.~Li, J.~Ruan, Z.~Xie, H.~Wang, and W.~Liu, ``{Investigation of the critical
  geometric characteristics of living human skulls utilising medical image
  analysis techniques},'' \emph{International Journal of Vehicle Safety},
  vol.~2, jan 2007.

\bibitem{Dempsey2017}
L.~A. Dempsey, M.~Persad, S.~Powell, D.~Chitnis, and J.~C. Hebden,
  ``\BIBforeignlanguage{eng}{{Geometrically complex 3D-printed phantoms for
  diffuse optical imaging}},'' \emph{\BIBforeignlanguage{eng}{Biomedical optics
  express}}, vol.~8, no.~3, pp. 1754--1762, feb 2017.

\bibitem{Bouchard2010}
J.-P. Bouchard, I.~Veilleux, R.~Jedidi, I.~Noiseux, M.~Fortin, and O.~Mermut,
  ``Reference optical phantoms for diffuse optical spectroscopy. part 1 --
  error analysis of a time resolved transmittance characterization method,''
  \emph{Opt. Express}, vol.~18, no.~11, pp. 11\,495--11\,507, May 2010.

\bibitem{Comiskey1998}
\BIBentryALTinterwordspacing
B.~Comiskey, J.~D. Albert, H.~Yoshizawa, and J.~Jacobson, ``{An electrophoretic
  ink for all-printed reflective electronic displays},'' \emph{Nature}, vol.
  394, no. 6690, pp. 253--255, 1998. [Online]. Available:
  \url{https://doi.org/10.1038/28349}
\BIBentrySTDinterwordspacing

\end{thebibliography}

\ifpeerreview \else





\begin{IEEEbiography}[{\includegraphics[width=1in,height=1.25in,clip,keepaspectratio]{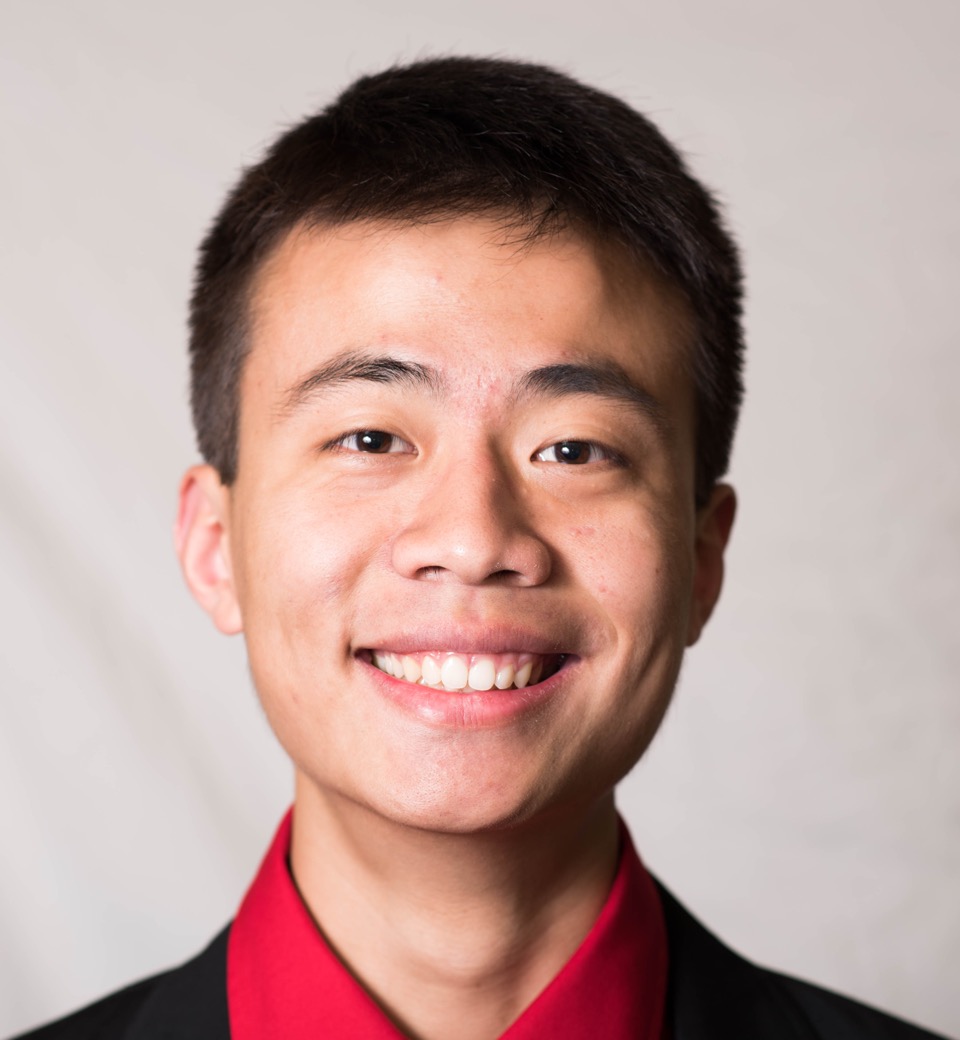}}]{Yongyi Zhao} Yongyi Zhao is a Ph.D student in the Electrical and Computer Engineering (ECE) department at Rice University. His research focuses on algorithms for imaging through densely scattering media. Before joining Rice, he received his B.S. degree in ECE from Carnegie Mellon University.
\end{IEEEbiography}
\begin{IEEEbiography}[{\includegraphics[width=1in,height=1.25in,clip,keepaspectratio]{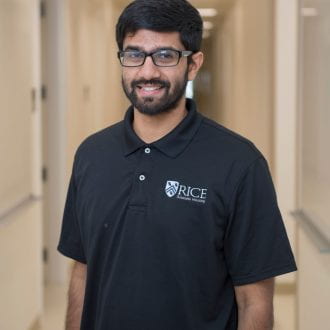}}]{Ankit Raghuram}
Ankit Raghuram received his B.S. degree in Biomedical Engineering from the Georgia Institute of Technology in 2015 and his M.S degree in ECE from Rice University in 2020. He is currently a Ph.D student at Rice University, and his interests include biomedical imaging, microscopy, diffuse imaging, time-of-flight, and computational imaging.
\end{IEEEbiography}
\begin{IEEEbiography}[{\includegraphics[width=1in,height=1.25in,clip,keepaspectratio]{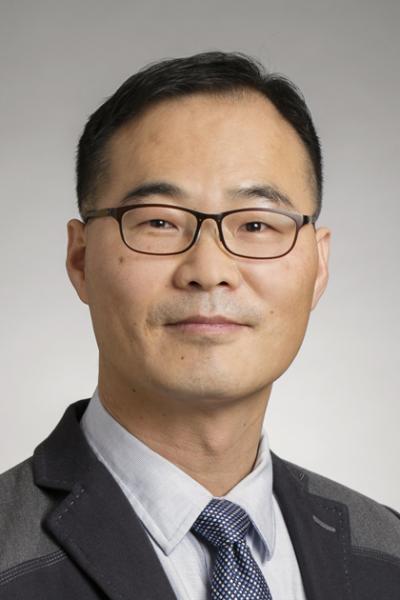}}]{Hyun K. Kim}
Dr. Hyun K. Kim is an Associate Professor of Radiology (Physics) at Columbia University Medical Center. Dr. Kim earned his PhD. degree in Mechanical Engineering at Korea Advanced Institute of Sciences and Technology (KAIST) in 2004. Dr. Kim has expertise in the development of mathematical and physical models of transport-related biological systems, which includes characterization of light-tissue interaction, 3D imaging of physiological parameters, and 3D molecular imaging.
\end{IEEEbiography}
\begin{IEEEbiography}[{\includegraphics[width=1in,height=1.25in,clip,keepaspectratio]{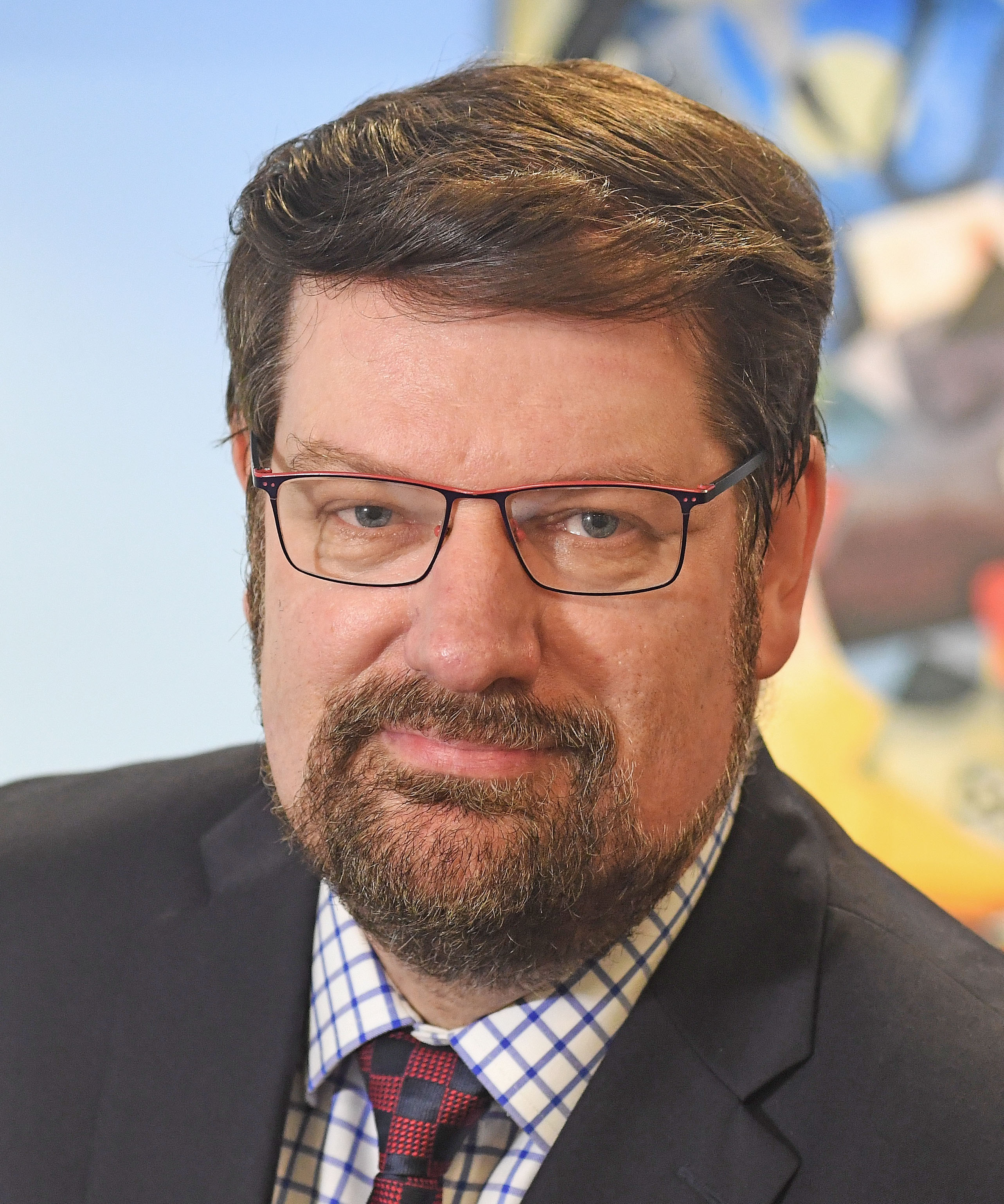}}]{Andreas H. Hielscher}
Dr. Andreas H. Hielscher received his PhD degree in Electrical and Computer Engineering from Rice University, Houston, Texas.  After a postdoctoral fellowship at the Los Alamos National Laboratory in New Mexico, he moved to Columbia University, in New York City, in 2001.  In summer 2020, he moved across town to head the newly formed Department of Biomedical Engineering at New York University. His work focuses on the development of  optical tomographic imaging systems that his team applies to brain imaging, diagnose and monitor arthritis, vascular diseases, breast cancer.
\end{IEEEbiography}
\begin{IEEEbiography}[{\includegraphics[width=1in,height=1.25in,clip,keepaspectratio]{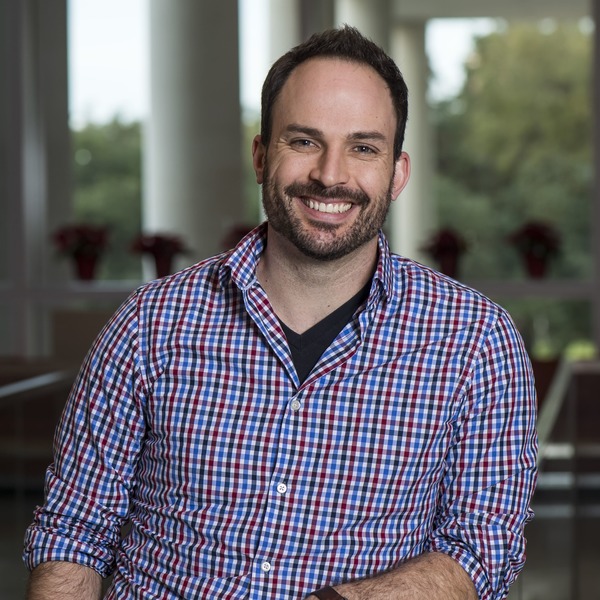}}]{Jacob T. Robinson}
Dr. Jacob Robinson is an Associate professor of Electrical and Computer Engineering at Rice University and a member of the Rice Neuroengineering Initiative. Dr. Robinson completed his Ph.D. in Applied Physics from Cornell University in 2008. He then joined Professor Hongkun Park's research group in the Chemistry and Chemical Biology Department at Harvard University. Currently at Rice University, his research interests include nanoelectronic, nanophotonic and nanomagnetic technologies to manipulate and measure brain activity.
\end{IEEEbiography}
\begin{IEEEbiography}[{\includegraphics[width=1in,height=1.25in,clip,keepaspectratio]{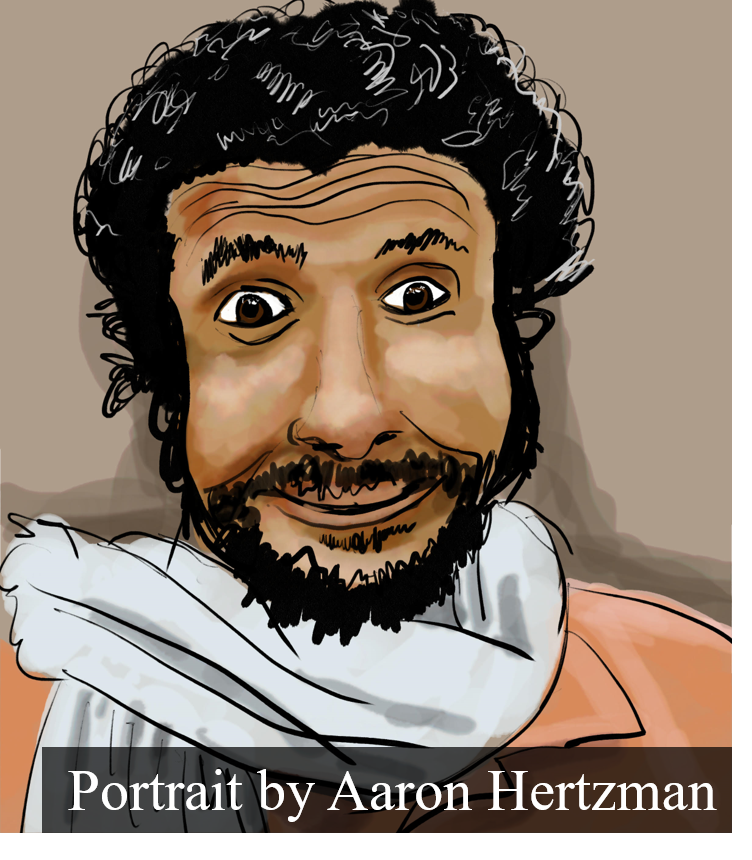}}]{Ashok Veeraraghavan}
Dr. Ashok Veeraraghavan is a Professor of Electrical and Computer Engineering at Rice University. Dr. Veeraraghavan received his Ph.D. in Electrical and Computer Engineering from the University of Maryland in 2008. He then worked as a research scientist at Mitsubishi Electronics Research Laboratory until 2011, when he joined the faculty at Rice University. His research interests are broadly in the areas of  imaging, vision, and machine learning and their applications to health and neuroengineering. 
\end{IEEEbiography}
\vfill


\fi

\end{document}